\documentclass[traditabstract,longauth]{aa}
\usepackage{amssymb}
\usepackage{mathptmx}
\usepackage{natbib}
\usepackage[]{graphicx}

\begin{document}

\title{Defying jet-gas alignment in two radio galaxies at z$\sim$2
 with extended light profiles: Similarities to brightest cluster
 galaxies 
\thanks{Based on observations carried out with the Very Large
Telescope of ESO under Program IDs 079.A-0617, 084.A-0324, 
085.A-0897, and 090.A-0614.}}
\author{C.~Collet\inst{1}, N.~P.~H.~Nesvadba\inst{1,2},
  C.~De~Breuck\inst{3}, M.~D.~Lehnert\inst{4}, P.~Best\inst{5},
  J.~J.~Bryant\inst{6,7,8}, D.~Dicken\inst{1,9}, H.~Johnston\inst{6},
  R.~Hunstead\inst{6} and D.~Wylezalek\inst{3,10}}
\institute{Institut d'Astrophysique Spatiale, CNRS, Centre Universitaire
  d'Orsay, Bat. 120-121, 91405 Orsay, France 
\and 
email: nicole.nesvadba@ias.u-psud.fr
\and
European Southern Observatory, Karl-Schwarzschild Strasse, Garching bei M\"unchen, Germany
\and
Institut d'Astrophysique de Paris, UMR 7095 CNRS, Universit\'e Pierre et Marie Curie, 98 bis, 
 boulevard Arago, 75014 Paris, France
\and
SUPA, Institute for Astronomy, Royal Observatory of Edinburgh, Blackford Hill, Edinburgh EH9 3HJ, UK
\and
Sydney Institute for Astronomy (SIfA), School of Physics, The University of Sydney, NSW 2006, Australia 
\and
Australian Astronomical Observatory, PO Box 915, North Ryde, NSW1670, Australia 
\and
ARC Centre of Excellence for All-sky Astrophysics (CAASTRO), Australia
\and
Laboratoire AIM Paris-Saclay, CEA/DSM/Irfu, Orme des Merisiers, Bat 709, 91191 Gif sur Yvette, France
\and
Department of Physics and Astronomy, Johns Hopkins University, 3400 N. Charles St, Baltimore, MD 21218, USA
}
\titlerunning{Defying the alignment effect in high-z radio galaxies}
\authorrunning{Collet et al.}  \date{Received / Accepted }

\date{Accepted. Received ; in original form }

\abstract{
We report the detection of extended warm ionized gas in two powerful
high-redshift radio galaxies, NVSS~J210626-314003 at z$=$2.10 and
TXS~2353-003 at z$=$1.49, that does not appear to be associated with
the radio jets. This is contrary to what would be expected from the alignment
effect, a characteristic feature of distant, powerful radio galaxies
at z$\ge$0.6. The gas also has 
smaller velocity gradients and line widths than most other high-z 
  radio galaxies with similar data. Both galaxies are part of a systematic study of 50
high-redshift radio galaxies with SINFONI, and are the only two
that are characterized by the {\it \emph{presence}} of high
surface-brightness gas not associated with the jet axis and by the
{\it \emph{absence}} of such gas aligned with the jet. Both galaxies are spatially
resolved with ISAAC broadband imaging covering the rest-frame R band, and have 
extended wings that cannot be attributed to line
contamination. We argue that the gas and stellar properties of these
galaxies are more akin to gas-rich brightest cluster galaxies in
cool-core clusters than the general population of high-redshift radio
galaxies at z$\ga$2. In support of this interpretation, one of our
sources, TXS~2353-003, for which we have H$\alpha$ narrowband
imaging, is associated with an overdensity of candidate H$\alpha$
emitters by a factor of $\sim$8 relative to the field at z$=$1.5. We discuss
possible scenarios of the evolutionary state of these galaxies and the
nature of their emission line gas within the context of cyclical AGN
feedback.}

\keywords{
galaxies: formation, galaxies: high-redshift, quasars: emission lines,
galaxies: kinematics and dynamics}

\maketitle

\section{Introduction}
\label{sec:introduction}
Powerful radio galaxies at redshifts z$\ge$1-2 (HzRGs) are
particularly massive \citep[several $10^{11}$ M$_{\odot}$,
  e.g.,][]{seymour07, debreuck10}, intensely star-forming
\citep[on the order of  1000 M$_{\odot}$ yr$^{-1}$][]{archibald01, reuland04,
  seymour12, ivison12, barthel12, drouart14}, relatively evolved
\citep[][]{nesvadba11a} galaxies. 
Many are surrounded by overdensities
of satellite galaxies, making them interesting tracers of overdense
regions in the early Universe that may ultimately evolve into
massive, rich galaxy clusters \citep[e.g.,][]{chambers96,lefevre96,
  Kurk2004a, venemans07, galametz12, wylezalek13}. Their co-moving number
density corresponds to that of massive clusters, when corrected for
duty-cycle effects \citep[][]{venemans07,
  Nesvadba2008}. Particularly detailed studies of individual HzRGs
suggest that at least one subset of this population could be the
progenitors of the central galaxies in massive clusters in the nearby
Universe \citep[e.g.,][]{hatch09}. They follow the low-redshift
  Kormendy relation of powerful radio and brightest cluster galaxies
  when allowing for passive luminosity evolution
  \citep[][]{targett11}.  Many HzRGs are themselves gas rich
\citep[e.g.,][]{vanOjik1997, nesvadba08, ogle12, emonts14}, and while
they have no well established, virialized halos of hot gas
\citep[][]{overzier05}, HzRGs do have extended (few 100 kpc)
reservoirs of diffuse warm ionized gas \citep[][]{villar03} with
embedded clouds or filaments of neutral \citep[][]{vanOjik1997, adams09}
and molecular gas \citep[][]{nesvadba09, debreuck03, emonts14}. As
suggested by, e.g., \citet[][]{villar03} these could be the vestiges
of the gaseous reservoirs from which the brightest cluster and other
galaxies formed.

The central, high surface-brightness regions of HzRGs are well
  fitted with de Vaucouleurs profiles with relatively large radii,
  R$\sim$8~kpc \citep{targett11}, but extended, often irregular line
and continuum emission in HzRGs is found preferentially along
the radio jet axis \citep[e.g.,][]{cimatti93}. The reasons for this
alignment effect are still not fully understood. Polarimetry suggests
that light scattering on dust grains could be the origin of at least
some of the extended continuum light; other authors favor
jet-triggered star formation \citep[][]{chambers88, klamer05} as
the cause of the extended continuum emission.

The kinematics in the extended gas around HzRGs are often strongly
perturbed \citep[e.g.,][]{baum00, villar03}. SINFONI imaging
spectroscopy showed this is   the case in
individual regions of these galaxies, and throughout their extended
emission-line gas \citep[][]{nesvadba06, nesvadba08}. Warm ionized
gas seems to be a major component of the ISM in these galaxies, with
ionized gas masses of up to $\gtrsim 10^{10}$ M$_{\odot}$ in some
cases \citep[][]{nesvadba08, nesvadba06}, comparable to the molecular
gas masses found in some HzRGs
\citep[e.g.,][]{emonts14,nesvadba09,debreuck05}. Broad line widths and
abrupt velocity offsets of up to 2000 km s$^{-1}$ found in this gas
are not reconcilable with gravitational motion. Instead they suggest that a
small part of the kinetic energy release from the AGN, through powerful
radio jets, accelerates large fractions of the ISM in these galaxies to
velocities well above the escape velocity from the underlying
dark-matter halo, thereby driving winds and removing the fuel for
subsequent star formation. This scenario is broadly supported by
hydrodynamic models showing that jets may indeed accelerate relatively
dense gas to the observed velocities of a few 100 km s$^{-1}$
\citep[][]{Wagner2012}.

Using VLT near-infrared imaging spectroscopy we have carried
out a large survey of the rest-frame optical line emission of the warm
ionized gas in a set of 50 radio-selected HzRGs at
z$\ga$2 spanning three orders of magnitude in radio power from
a few times $10^{26}$ to a few times $10^{29}$ W Hz$^{-1}$ at 1.4 GHz in the
 rest frame, and more than two orders of magnitude in radio size,
 from 1~kpc to about 300~kpc (Collet et al. 2015, submitted, Nesvadba
 et al. 2015, in prep.). Emission-line regions are well aligned with
the radio jet axis in all but two galaxies with extended line
emission, which in addition to kinematic, energy, and timing arguments
suggests that this gas represents outflows of warm ionized gas driven
by the overpressurized cocoon of hot plasma inflated by the radio
source. These galaxies are  described in greater detail in \citet{nesvadba06,
  nesvadba07, nesvadba08, nesvadba11a} and Collet et al. (2015).

Here we discuss the two radio galaxies, TXS~2353$-$003 and
NVSS~J210626$-$314003, that do not follow these global trends. Their
extended regions of warm ionized gas with similar surface brightness
to the other sources show strong offsets in position angle to the
radio jet axis, contrary to what is expected from the alignment effect and
observed in the other galaxies within the SINFONI sample.  Gas
associated with the radio jet axis, if present at all, is not a
distinctive component compared to gas along other directions in our
SINFONI data.   Gas that is seen in projection to be associated
with the jet axis may or may not imply a physical association; however,  it is
clear that gas that is globally and significantly offset from the jet
axis as projected on the sky is not reconcilable with a scenario where
the gas is embedded in an overpressured cocoon inflated by the
on-going jet activity. In addition, the line widths are low compared
to the rest of our sample.  Line widths near the nucleus are higher,
and may be affected by the central radio source.

TXS~2353-003 and NVSS~J210626-314003 are among  the sources with the
largest radio sizes in our sample, with largest angular sizes
LAS$=$38.8\arcsec\ and 24.2\arcsec\ at 4.86 GHz \citep[][]{debreuck01}
and 2.368~GHz, respectively (corresponding to 310~kpc and 190~kpc at
their redshifts of z=1.49 and z=2.10, respectively). Both have very
regular centimeter radio morphologies (Collet et al. 2015, Nesvadba et
al. 2015 in prep.), and do not have bright radio cores. Both are
associated with overdensities seen in projection along nearby lines of
sight of distant (z$\ge$1.3) IRAC sources in the CARLA survey of
\citet{wylezalek13}. TXS~2353-003 is within the densest region of
their overall sample of 200 HzRGs. We highlight the similarities
between our galaxies and brightest cluster galaxies at low redshift,
including the extended emission-line gas, which resembles the warm
ionized gas in and around the central galaxies of cool-core clusters
in several respects.

\begin{figure}[t]
\begin{center}
\includegraphics[width=0.45\textwidth]{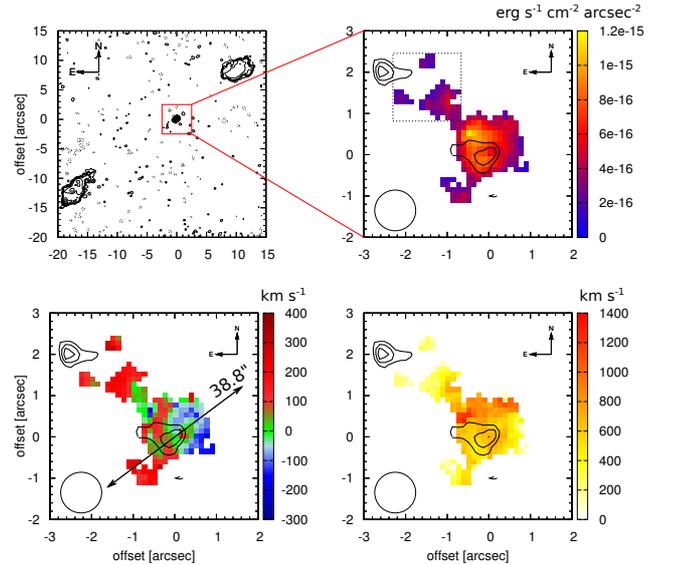}
\caption{Results of our observations of TXS~2353-003. 
{\it Top left:} 5.5~GHz contours from ATCA. 
{\it Clockwise from top right to bottom left:} SINFONI maps of
[OIII]$\lambda$5007 surface brightness, FWHM line widths, and relative
velocities.  The square in the top left panel shows the size of the SINFONI
maps compared to radio size. Black contours show the 5.5~GHz radio
continuum, and the dotted square in the top right panel illustrates
the region from which the bottom spectrum in
Fig.~\ref{fig:TXS2355spectra} was extracted. The arrow in the
 lower left panel shows the direction of the radio jet, and the
 number gives the jet size in arcsec. Ellipses in the lower left
 corner of each panel show the FWHM size of the point spread
 function.}
\label{fig:SINFONIobsTXS2355}
\end{center}
\end{figure}

We describe our SINFONI and ISAAC observations and data reduction in
\S\ref{sec:observations}, and discuss the results of these
observations in \S\ref{ssec:results:linemission} for the line emission
and in \S\ref{sec:contmorphologies} for the continuum, where we also
fit surface brightness profiles. In \S\ref{sec:environment} we present
our narrowband search for line emitters around TXS~2353-003 and argue
that this galaxy is surrounded by an overdensity of line emitters,
like many other HzRGs at somewhat greater redshifts. We compare these results with
the IRAC results obtained as part of the CARLA survey
\citep[][]{wylezalek13} in \S\ref{sec:bcgs}. In \S\ref{sec:nature} we
discuss several hypotheses regarding the nature of the emission-line
regions in these two galaxies, which, given their strong misalignment
relative to the jet axis, are unlikely to be directly associated with
an expanding cocoon driving an outflow. We summarize our results in
\S\ref{sec:summary}.  Throughout our analysis we adopt a flat
cosmology with H$_0$ = 70 km s$^{-1}$ Mpc$^{-1}$,
$\Omega_{\Lambda}$=0.7, $\Omega_{M}=0.3$. For NVSS~J210626-314003 at
z=2.1, this corresponds to D$_L=$ 16.6~Gpc, and for TXS~2353$-$003 at
z=1.49 to D$_L$=10.8~Gpc, respectively; 1\arcsec\ corresponds to
8.4~kpc for both galaxies.

\begin{figure}[t]
\begin{center}
\includegraphics[width=0.45\textwidth]{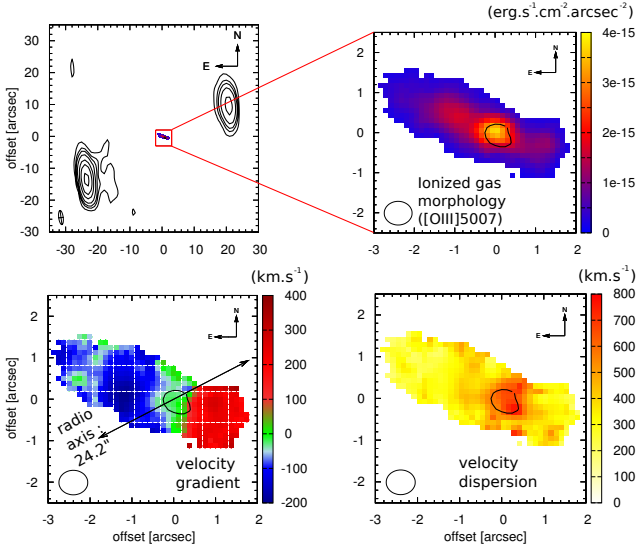}
\caption{Results of our observations of NVSS~J210626$-$314003. 
{\it Top left:} 5.5~GHz contours from ATCA. 
{\it Clockwise from top right to bottom left:} SINFONI maps of
[OIII]$\lambda$5007 surface brightness, FWHM line widths, and relative
velocities. The square in the top left shows the size of the SINFONI
maps compared to radio size. The black contour marks the continuum
position in our SINFONI cube. Ellipses in the lower left corner
  of each panel show the FWHM size of the point spread function.}
\label{fig:SINFONIobsNVSS2106}
\end{center}
\end{figure}

\section{Observations and data reduction}
\label{sec:observations}

\subsection{VLT/SINFONI imaging spectroscopy} 
\label{ssec:observations:sinfoni}

We observed both galaxies with the near-infrared imaging spectrograph
SINFONI on UT~4 of the Very Large Telescope (VLT) of ESO. The 
NVSS~J210626-314003 data were taken on 2009 October 12, and those
of TXS~2353-003 on 2010 August 13. The data were obtained as part
of a comprehensive observational program to study the emission-line
gas kinematics of 50 powerful radio galaxies at z$\ge$2. We used the
H$+$K band covering a spectral range of 1.45-2.45$\mu$m at a
spectral resolving power R$=$1500. The data were taken during good to
moderate observing conditions, with a typical point spread function of
1\arcsec$\times$1\arcsec\ along right ascension and declination, dominated by the seeing. We used the 250~mas pixel
scales with a field of view of 8\arcsec$\times$8\arcsec, covering a
field of 67~kpc$\times$67~kpc at z$=$2.

Total integration times were 145~min for TXS~2353-003 and 120~min for
NVSS~J210626-314003, and were obtained in one-hour
sequences of individual 300~s exposures. 
Our data reduction relied on a combination of IRAF and IDL
routines \citep[for details see, e.g., ][]{nesvadba11a, nesvadba11b}.
The absolute flux scale was measured by comparing our data with observations of
standard stars with known magnitudes from the 2MASS survey that
were observed at the end of each observing session. We also determined
the size of the seeing disk from these stars.

\subsection{VLT/ISAAC imaging}
\label{ssec:observations:isaac}

Both galaxies were also observed with ISAAC on UT~3 of the VLT between
2012 September 30 and 2012 October 2 through the
service-mode program 090.A-0614. TXS~2353$-$003 was observed in the
H band and through the narrowband  FeII filter (NB~1.64) with
FWHM$=0.025\mu$m (corresponding to a velocity range of $\Delta v=$4570
km s$^{-1}$), so we cover H$\alpha$ and [NII] at
z$=$1.49. NVSS~J210626-314003 was only observed in the Ks band. Both
broadband images cover roughly the rest-frame  R band.

With the NB~1.64 filter we obtained a total of 248~minutes of
on-source observing time of TXS~2353$-$003 split into 99 exposures
with individual observing times of 150~seconds, and grouped into
observing sequences  of one hour each. After
visual inspection we discarded 17 of these frames because of residuals
from a strongly saturated star. The on-source observing time was thus
205~minutes.

In the H band we obtained 29~exposures for TXS~2353-003 with
individual observing times of 72~seconds, corresponding to six detector
readouts after 12~seconds. We discarded one frame which contained
the trail of a satellite. In the K band for NVSS~J210626$-$314003, we
obtained 25~exposures of 90~seconds, corresponding to  six detector readouts after 15~seconds. Total observing times in the H and K bands were
therefore 35~minutes and 38~minutes for TXS~2353$-$003 and
NVSS~J210626$-$314003, respectively.

Data were dark-subtracted and flat-fielded, where we constructed the
sky flats directly from the science data. All frames were then
combined with the IRAF task {\verb xdimsum }. Flux calibration relied
on standard stars observed during the same night and at similar air
mass to the science data. We used the Starlink GAIA software
\citep[][]{draper14} to calculate the astrometry relative to 2MASS
reaching rms$=$0.07\arcsec\ for TXS~2353$-$003, and
rms$=$0.3\arcsec\ for NVSS~J210626$-$314003, respectively. We also
used 2MASS to verify our flux calibration, finding a good agreement
within $\la$10\%. The sizes of stars in the image suggest the seeing
was FWHM$=$0.5\arcsec\ in the H-band and NB~1.64 image of
TXS~2353$-$003, and of FWHM$=$0.4\arcsec\ in the Ks-band image of
NVSS~J210626-314003. The limiting rms surface brightness in these images
is 26.2 mag arcsec$^{-2}$ and 23.6~mag~arcsec$^{-2}$ in the narrow
and broadband image of TXS~2353$-$003, respectively. For the Ks-band
image of NVSS~J210626$-$314003, we find 23.7~mag arcsec$^{-2}$.

\subsection{Narrowband observations of TXS~2353$-$003}
\label{ssec:narrowband}
To construct a continuum-subtracted line image from the narrowband
image of TXS~2353-003, we approximate the underlying continuum in the
narrowband filter by a scaled version of the flux measured through
the broadband filter, taking into account the relative filter
bandwidths of the narrow and broadband filters, and their
transmission. Here we assume that the continuum is flat and uniform
across the H-band filter.

We measured the zero points in the narrowband filter from
standard stars. From the zero points in each filter, filter
bandwidths, and transmissions of the narrow and broadband filters,
we determine that the continuum flux density observed through the
narrowband image corresponds to about 7.2\% of the flux density
measured through the broadband image. We would therefore need to
scale the broadband image by this value before subtracting it from
the narrowband image to obtain the flux density from the emission
lines. When directly comparing the flux density of 12 stars in both
images, we empirically find that a very similar scaling factor of
7.6\% is best to minimize the residuals of the stellar continuum
after the subtraction, and this is the factor that we used.

The redshifted wavelength of H$\alpha$ in the radio galaxy is slightly
blueshifted relative to the wavelength of peak throughput of the
narrowband filter. Comparing this with the flux measured in the SINFONI
data cube, and assuming that the relative flux calibration between
both images is perfect, we find that we are missing between 20 and
25\% of the line flux in the continuum-subtracted line image. For
other H$\alpha$ emitters (\S\ref{sec:environment}), for which we have
no precise redshift estimates, we would first need to obtain
spectroscopy before we can make similar corrections.

To evaluate the completeness limit of our data, we added faint
artificial sources with the size of the seeing disk at sky positions
in each frame of the narrow- and broadband images
before re-reducing these data and extracting sources in the same way
as for the final scientific analysis. We used fluxes between 2 and
11 times  the rms in the central regions of the image. Adding the
artificial sources to the individual frames rather than the final
image helps to take misalignments between frames into account. We find
that we recover 90\% of sources with surface brightness above $1.5
\times 10^{-19}$~erg s$^{-1}$ cm$^{-2}$~\AA$^{-1}$ arcsec$^{-2}$ (or
above 23.6~mag$_{AB}$~arcsec$^{-2}$).

\section{Imaging spectroscopy of the warm ionized gas}
\label{ssec:results:linemission}
The SINFONI maps of NVSS~J210626-314003 and TXS~2353-003 are shown in
Figs.~\ref{fig:SINFONIobsTXS2355} and \ref{fig:SINFONIobsNVSS2106},
respectively.  In NVSS~J210626$-$314003 at z$=$2.104,
[OIII]$\lambda\lambda$4959,5007 and H$\beta$ fall into the H band, and
H$\alpha$, [NII]$\lambda\lambda6548,6583$, and
[SII]$\lambda\lambda6716,6731$ fall into the K band, both of which we cover
 with our data cubes. In TXS~2353$-$003, we identify H$\alpha$ and
[NII]$\lambda\lambda6548,6583$ at wavelengths corresponding to a
redshift of z$=$1.49. At this redshift,
[OIII]$\lambda\lambda4959,5007$ and H$\beta$ fall at redshifts below
the lower wavelength cutoff of our grating, $\lambda_{\rm
 min}=1.45\ \mu$m.

Our redshift measurement for TXS~2353-003 does not agree with the
previous redshift of z$=$2.59 measured with longslit spectroscopy,
which was thought to have detected Ly$\alpha$ in the observed optical wavelength
range \citep{debreuck01}, but was in hindsight probably affected by a
Cosmic ray. We did not identify any possible emission line consistent
with the previous redshift z$=$2.59, although
[OIII]$\lambda\lambda$4959,5007 and H$\beta$ should fall into the
H band, and [NII]$\lambda\lambda$6548,6583, and H$\alpha$ should
fall into the K band at that redshift. The rest-frame UV spectrum
shows no other line candidate consistent with z$=$2.59; the value for Ly$\alpha$ at
the redshift $z=1.49$ we measure here falls well below the UV
atmospheric cutoff.

Both galaxies have bright, spatially well-resolved line emission. We
fitted the extended line emission in both galaxies with single
Gaussian profiles. Spectra were extracted from small apertures of
0.4\arcsec$\times$0.4\arcsec, which helps to reduce the strongest
pixel-to-pixel noise while safely oversampling the seeing disk so that
no spatial information is lost. The data were also smoothed in the
spectral direction by three pixels, without loss of spectral
resolution. Our fitting routine interpolates over wavelength ranges
dominated by bright night-sky line residuals, and we only map spatial
pixels where a line was detected at a S/N $>$3.

Line emission extends beyond the bright continuum seen in the SINFONI
cubes (but not in the ISAAC broadband image which we will discuss
later in more detail and show in Figs.~\ref{fig:txs2353cont}
and~\ref{fig:nvss2106cont}, which is deeper in the continuum). The
sizes of the extended emission-line regions of NVSS J210626-314003 and
TXS~2353$-$003 are 5.3\arcsec$\times$2.0\arcsec\
(44~kpc$\times$17~kpc) and 2.5\arcsec$\times$2.0\arcsec
(21~kpc$\times$17~kpc) along the major and minor axis, respectively,
and down to surface-brightness limits of $4.3\times 10^{-17}$ erg
s$^{-1}$ cm$^{-2}$ arcsec$^{-2}$ and $6.2\times 10^{-17}$ erg s$^{-1}$
cm$^{-2}$ arcsec$^{-2}$.

\begin{figure}
\begin{center}
\includegraphics[width=0.48\textwidth]{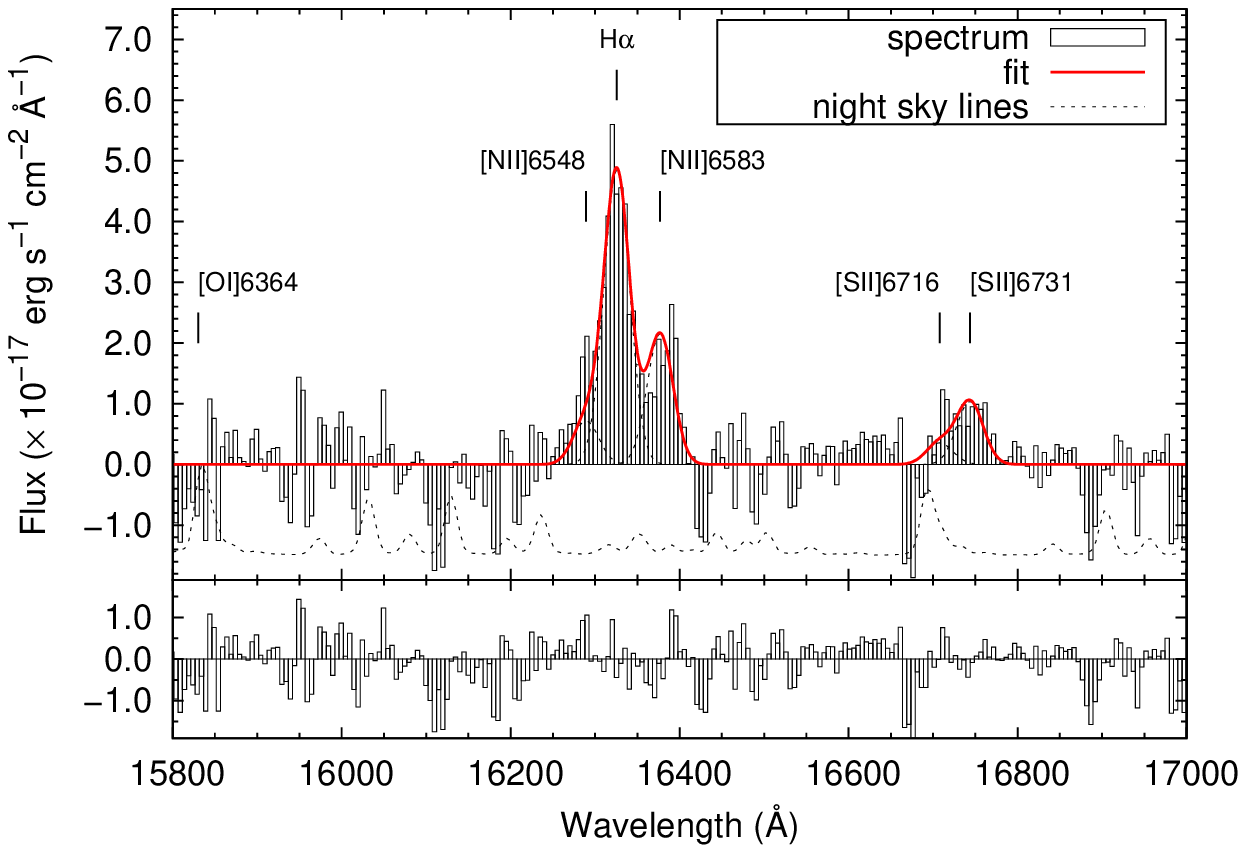}
\includegraphics[width=0.48\textwidth]{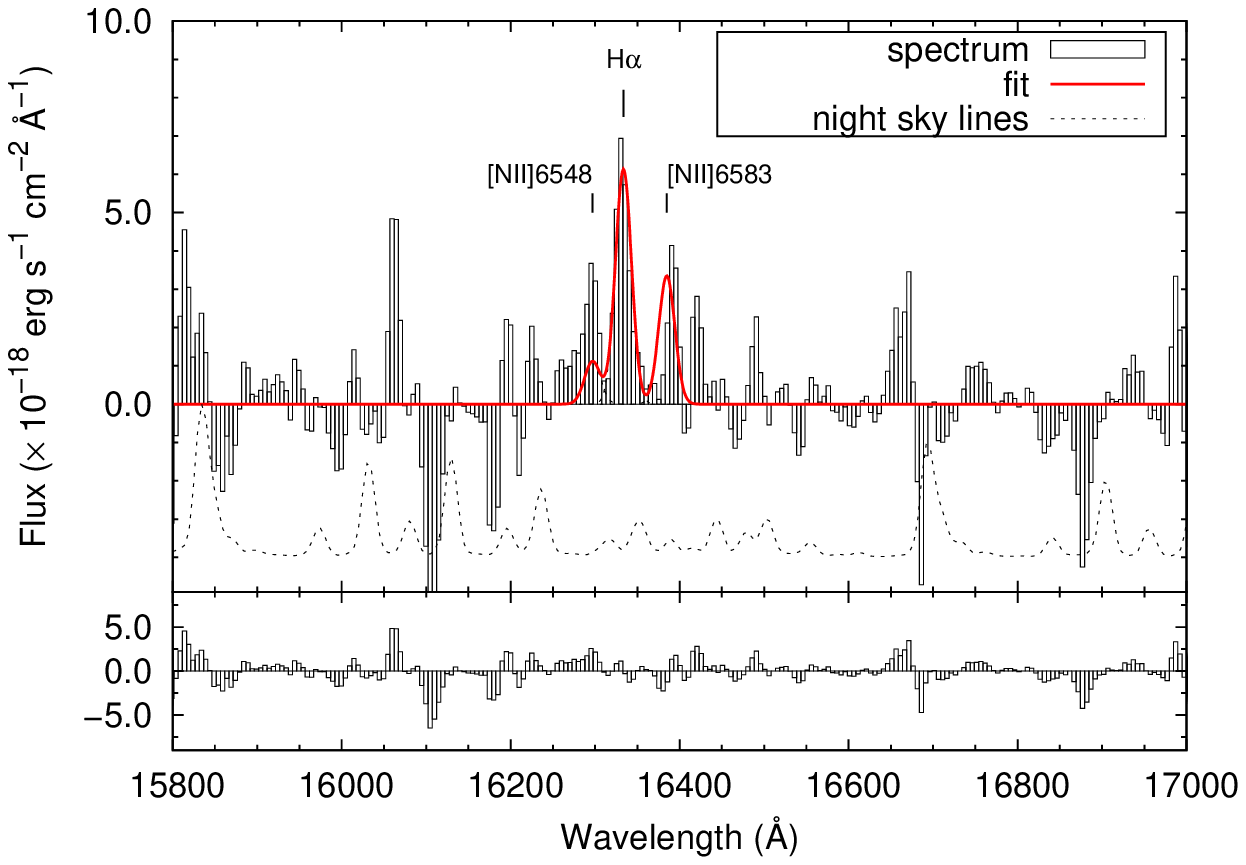}
\end{center}
\caption{{\it Top}:
Spectrum of TXS~2353$-$003, integrated over all spatial pixels
 where the H$\alpha$ emission from the galaxy is detected at $\ge
 3\sigma$, and corrected for velocity shifts.
{\it Bottom}: Integrated
spectrum of the northeastern part of the emission-line region of
TXS~2353$-$003, defined by the dotted square in
Fig.~\ref{fig:SINFONIobsTXS2355}.}
\label{fig:TXS2355spectra}
\end{figure}

In NVSS~J210626$-$314003 the emission-line region is roughly centered
on the continuum peak. In TXS~2353$-$003, the gas appears more
lopsided. This is also seen in the continuum-subtracted ISAAC
narrowband image (source $a$ in Fig.~\ref{fig:thumbnailsHAE}). The
morphology of the ionized gas in both observations is fairly
consistent: a bright emission-line region associated with the radio
galaxy, more extended toward the southeast and northwest. The integrated
spectrum extracted from this fainter region, shown as a dashed region in
Fig.~\ref{fig:SINFONIobsTXS2355}, clearly shows H$\alpha$ and
[NII]$\lambda$6583 (Fig.~\ref{fig:TXS2355spectra}). Fluxes and other
line properties are listed for both galaxies in
Table~\ref{tab:propertiesUnalignedGalaxies}. In TXS~2353$-$003, we
note a faint continuum emitter along the direction of extended gas,
but at a larger distance. In NVSS~J210626$-$314003 no further
continuum source is observed along the major axis of the emission-line
region.

In the SINFONI cube, NVSS~210626-3134003 shows a smooth velocity
gradient of about 600~km s$^{-1}$ over 25~kpc. In
TXS~2353$-$003 we observe a more irregular velocity pattern, which
could be at least in part due to the lower signal-to-noise ratio and
irregular line profiles. In NVSS~J210626-3134003, where we observe
[OIII]$\lambda$5007 and H$\alpha$ at good signal-to-noise ratios, the
gas kinematics and morphologies are similar in both lines.

Both galaxies have broad line widths with FWHM$\sim$800~km~s$^{-1}$
around the continuum peak which we associate with the central regions
of the galaxies. These lines are broader than those of star-forming
galaxies at similar redshifts, including the most massive, intensely
star-forming dusty galaxies such as submillimeter galaxies
\citep[e.g.,][]{swinbank06}, and are consistent with those found in other
powerful radio galaxies at similar redshifts
\citep[e.g.,][]{nesvadba08, collet14}. Following the arguments given
in these studies, we consider the broad line widths as evidence that
the gas in the central regions of our sources is stirred up by the
transfer of kinetic energy from the radio source.

The gas kinematics at larger radii are more quiescent in both
sources. Line widths in the extended gas of NVSS~J210626-314003 and
TXS~2353-003 are FWHM$=$300-400 km s$^{-1}$
(Figs.~ \ref{fig:SINFONIobsTXS2355} and \ref{fig:SINFONIobsNVSS2106}),
more akin to those observed in a quasar illumination cone at z$\sim$2
\citep[][]{lehnert09} and Ly$\alpha$ blobs without a powerful central
radio source \citep[e.g.,][]{wilman05,overzier13}, than to powerful
radio galaxies, including the central regions of our two sources here.

\begin{figure}
\begin{center}
\includegraphics[width=0.45\textwidth]{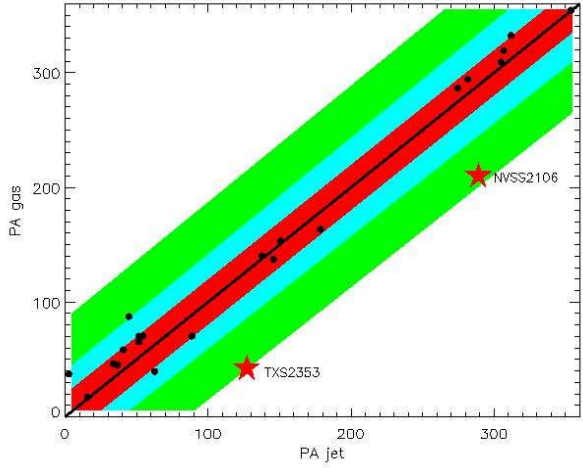}
\caption{Position angle (north through east) of the gas (ordinate) as
  a function of the position angle of the radio jet for our two
  sources. The outer boundaries of the red, blue, and green stripes
  indicate offsets of 20$^\circ$, 45$^\circ$, and 90$^\circ$,
  respectively. Red stars indicate our two sources discussed here,
  small black dots show the remaining HzRGs with SINFONI data and well-determined position angles (Nesvadba et al. 2015, in prep.).}
\label{fig:illustrationParticularityMisaligned}
\end{center}
\end{figure}

A marked difference to other HzRGs with similar data is that most of
the emitting gas is at large position angles from the radio jet
axis. Typically, the jet axis and the semi-major axis of the gas are
aligned within about 20$^\circ$-30$^\circ$
(Fig.~\ref{fig:illustrationParticularityMisaligned}, see
  also Collet et al. 2015, Nesvadba et al. 2015 in prep.). Given
the size of the minor axis of the extended line emission of up to
$\sim 15-20$ kpc and beam smearing effects in the radio and
near-infrared, this range is on the order of  what would be expected if gas
and jet were associated with each other, for example, if the jet
inflated a hot bubble of overpressurized gas like in the cocoon model
\citep[][]{begelman89}. In NVSS~J210626$-$314003 and TXS~2353$-$003;
however, the relative position angle between extended gas and radio
jet axis is much larger, 60$^\circ$ and 90$^\circ$, respectively, and
the jet is not embedded within the emission line gas, as expected from
a cocoon inflated by the
jet. Figure~\ref{fig:illustrationParticularityMisaligned} shows that
both galaxies clearly stand out from the overall sample of HzRGs with
SINFONI observations.

\subsection{Line diagnostics}

\begin{figure}
\includegraphics[width=0.45\textwidth]{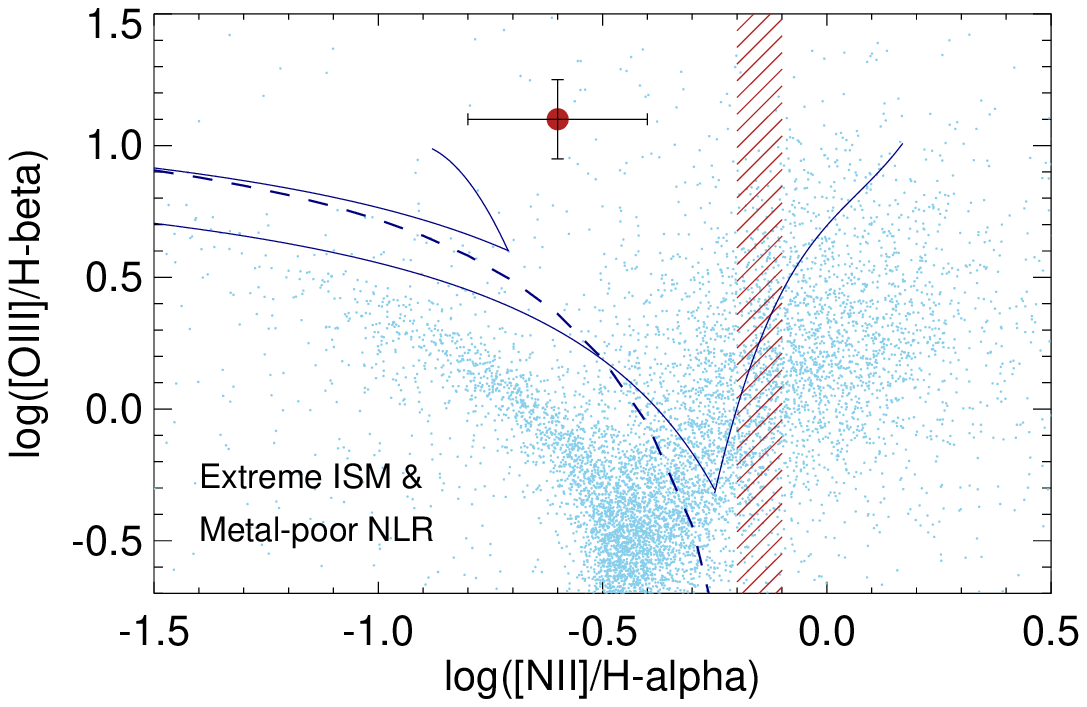}
\caption{Diagnostic diagram showing the ratio of [OIII]/H$\beta$
as a function of [NII]/H$\alpha$ line ratio \citep[][]{baldwin81,
VO87}. The red dot shows the position of
NVSS~J210626-314003. The red hatched area shows the region spanned
by the [NII]/H$\alpha$ ratio in the off-nucleus gas in
TXS~2353-003.  Light blue dots indicate the distribution of line
ratios for galaxies in the local Universe \citep[from SDSS;
 see][]{kauffmann03}, and the dashed blue line shows the
diagnostic of \citet{kewley06} to separate HII regions and
star-forming galaxies (below) from AGN (above). The solid blue
lines indicate how the distribution of line ratios for both SF
galaxies and AGN are expected to evolve out to high redshift
\citep[][their model 4 with metal-poor narrow-line
regions]{kewley13}: the two HzRGs are clearly consistent with
the predictions for high redshift AGN.}
\label{fig:bpt}
\end{figure}

The ratios of bright optical line emission provide important
constraints on the ionizing source and other gas properties in the
extended emission-line regions of galaxies
\citep[][]{baldwin81,VO87,kewley06}. \citet{collet14} discussed the
emission-line ratios of NVSS~J2106-314003 in the context of their
overall sample of moderately powerful HzRGs, finding [NII]/H$\alpha$
and [OIII]/H$\beta$ ratios that  correspond to  neither  the classical
starburst nor the AGN branch. They investigate several potential reasons for
this, including low metallicites, relatively diffuse gas clouds, high
gas pressures, or a mixture of heating from shocks and
photoionization. In TXS~2353-003, we only have [NII] and H$\alpha$
measured with a fairly high ratio of [NII]/H$\alpha$=0.7 in the
extended gas (Fig.~\ref{fig:TXS2355spectra}), which  alone
is sufficient to safely attribute the line emission to either shocks or
photoionization from the AGN (Fig.~\ref{fig:bpt}). Off-axis gas not
aligned with the radio jet axis and out to a galactocentric radius of
65~kpc in the radio galaxy 3C265 at z=0.8 may indeed be powered by
shocks, perhaps with additional photoionization from soft X-ray
emission from the precursor of the shock \citep{solorzano02}. However,
as argued by \citet{leTiran2011b} and \citet{Nesvadba2006}, it is
difficult to reach high H$\alpha$ surface brightnesses as observed
with shocks. The same is the case for shock-heated, likely infalling
clouds in the cluster central galaxy NGC~4696 \citep[][]{Farage2010}.

Like  TXS~2353-003, NVSS~J210626-314003 also falls outside
the star formation branch; however, it also falls outside the sequence
formed by low-redshift AGN in the SDSS. The line ratios are consistent
with low-metallicity gas photoionized by an AGN, as expected in
the models of \citet{groves06} and \citet{kewley13}, and as shown with 
the solid blue lines in Fig.~\ref{fig:bpt}. These lines show the range
of line ratios expected for galaxies at z=1.5 with intense
star formation  typical for galaxies at these redshifts, and AGN
with metal-poor narrow line regions and a metal enrichment history predicted
by cosmological models of galaxy evolution \citep[][]{dave11}.  These
conditions are referred to as ``model 4'' by
\citet[][]{kewley13}. J-band spectroscopy would be required to
investigate whether TXS~2355-003 falls onto the low-redshift AGN
branch or above.

\section{Continuum morphologies}
\label{sec:contmorphologies}

\subsection{Line-free continuum images from SINFONI}
We also constructed line-free continuum images from our SINFONI cubes
(shown as contours in
Figs.~\ref{fig:SINFONIobsTXS2355}~and~\ref{fig:SINFONIobsNVSS2106}) for
both galaxies by collapsing the data cubes along the spectral
direction over all wavelengths that are free from emission lines and
prominent night-sky line residuals. We consider the continuum peak in
the cube to be the location of the galaxy nucleus, which may be
dominated either by stellar light or the AGN.  At a projected distance
of about 3\arcsec\ (24~kpc) northeast of TXS~2353-003, we find
another, fainter H-band source. We do not detect any line emission
from this galaxy, so we cannot determine whether it is physically
associated with the radio galaxy, or an interloper along the line of
sight. NVSS~J210626$-$314003 appears to be an isolated source,
  except for two much fainter sources at 1.7\arcsec\ and
  3.9\arcsec\ toward the southeast (corresponding to projected
  separations of 14.3~kpc and 32.8~kpc at z$=$2.1,
  respectively), which however are not aligned with the extended
  emission-line regions. We do not detect any line emission from the
  nearer source, and the more distant source falls outside the field
  of view of SINFONI. It is therefore not clear if they are associated
  with the radio galaxy. For TXS~2353-003, for which we do not have
  broadband K-band imaging, we measure a K-band magnitude of
  18.1$\pm$0.2~mag from our SINFONI data cube within a
  3\arcsec\ aperture.

\subsection{ISAAC broadband imaging}
\label{ssec:results:continuum}

\begin{figure}
\begin{center}
\includegraphics[width=0.48\textwidth]{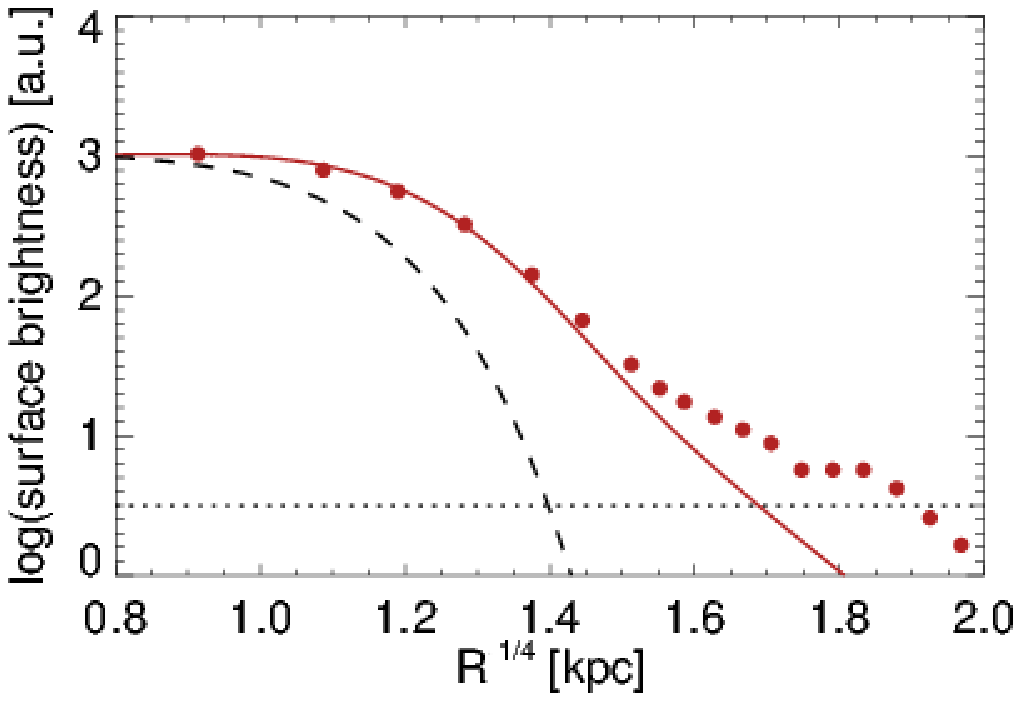}
\includegraphics[width=0.48\textwidth]{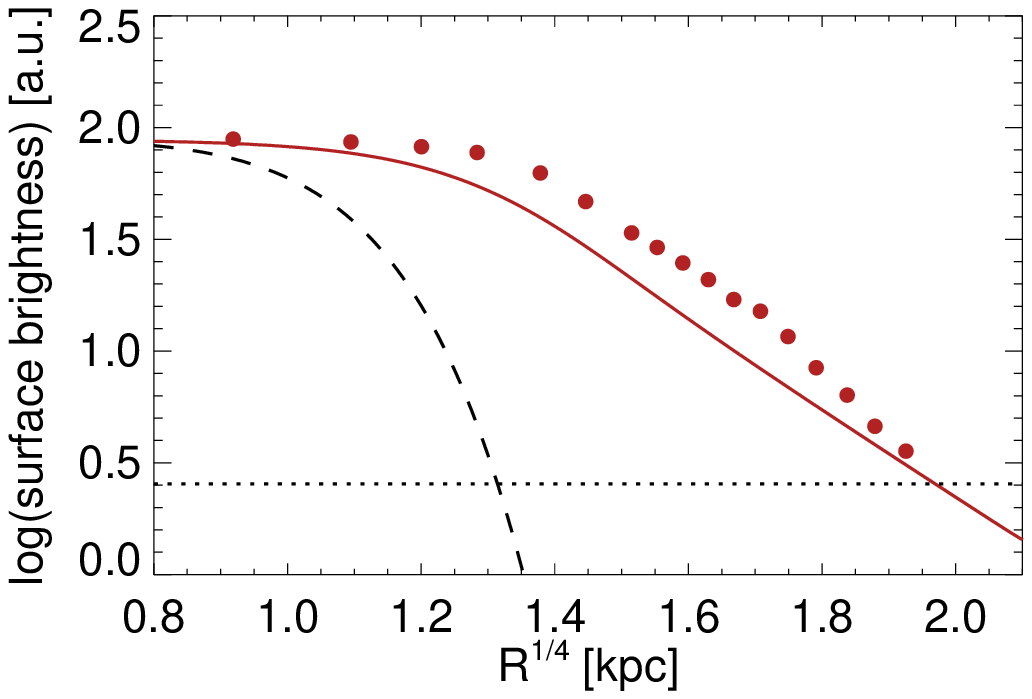}
\caption{Surface-brightness profiles of NVSS~J210626-314003 ({\it
    top}) and TXS~2353$-$003 ({\it bottom}) extracted from ISAAC
  H-band and Ks-band images, respectively. Red dots show the measured
  azimuthally averaged surface brightnesses as a function of
  R$^{1/4}$. Red lines show a Sersic fit with n$=$4, corresponding
  to the de Vaucouleurs law for classical elliptical galaxies, and
  convolved with the point spread function in each image. The dashed
  black lines show the size of the seeing disk. Dotted horizonal lines
  show the 3$\sigma$ limiting surface brightness of our images. Excess
  flux at large radii compared to the R$^{1/4}$ law is characteristic for
  brightest cluster galaxies in evolved galaxy clusters
  \citep[e.g.,][]{Schombert1987}.}
\label{fig:luminosityProfileChi2map_TXSB2353-003}
\end{center}
\end{figure}

In Figs.~\ref{fig:txs2353cont} and ~\ref{fig:nvss2106cont}, 
we have already shown the
rest-frame R-band morphologies of TXS~2353$-$003 and NVSS~J210626-314003
observed in the H and Ks band, respectively. Both galaxies are clearly
spatially resolved, even with ground-based, seeing-limited
images. This is best seen in
Fig.~\ref{fig:luminosityProfileChi2map_TXSB2353-003} which shows the
azimuthally averaged surface-brightness profiles of both sources
compared to the profiles of nearby stars taken from the same images,
showing a clear excess of continuum emission out to large radii
in both galaxies.

Both galaxies have a bright central region (which is more regular in
NVSS~J210626-314003 than in TXS~2353$-$003) surrounded by extended,
low-surface-brightness halos. This is remarkable, given that the
general population of passively evolving early-type galaxies at high
redshift tends to be more compact than their local analogs
\citep[e.g.,][]{Daddi2005, vanDokkum2008, Schawinski2011,
Whitaker2012}. Furthermore, although about 1/3 of HzRGs are still
actively star-forming \citep[][]{drouart14}, their molecular gas
reservoirs of a few $10^{10}$ M$_{\odot}$ \citep[e.g.,][]{emonts14}
compared to few $10^{11}$ M$_\odot$ of stellar mass
\citep[][]{seymour07, debreuck10} suggest that most of their stellar
mass growth has already been completed. \citet{targett11} found
extended broadband emission when stacking 13 HzRGs at redshifts
between z=1.5 and 2.0 \citep[as did][at slightly lower
redshifts]{best98}. However, only \citet{hatch09} have so far reported
a qualitatively similar surface-brightness profile in an individual
radio galaxy, MRC~1138$-$262 at z=2.16. Previous studies, e.g.,
\citet{pentericci01}, have drawn attention to the irregularity of
broadband morphologies in many cases, which they attributed to the
presence of on-going mergers.

\subsection{Surface brightness profiles}
\label{ssec:surfbrightprofile}

Both galaxies are spatially resolved into four or five resolution
elements in the ISAAC images, which enables us to do a basic analysis
of their surface-brightness profiles shown in
Fig.~\ref{fig:luminosityProfileChi2map_TXSB2353-003}. The goal of this
analysis is to demonstrate that the extended low-surface brightness
regions seen in the broadband images are not just the faint
extensions of the surface-brightness profiles of the central regions,
but separate components of the light profiles. We are
mainly interested in whether (and over what radii) the profiles are consistent with the
$r^{1/4}$ or de Vaucouleurs profile typical of early-type galaxies,
corresponding to a Sersic profile with index n$=$4
\citep[][]{Sersic1963, Sersic1968}.

The profiles shown in
Fig.~\ref{fig:luminosityProfileChi2map_TXSB2353-003} were obtained
with the IRAF task \verb+Ellipse+ \citep[][]{freudling93}, measuring
the surface brightness in circumnuclear elliptical annuli with radii
increasing in steps of 0.1\arcsec. \verb+Ellipse+ fits the ellipticity
of these annuli, finding between 0.05 and 0.4 in our case. Solid red
lines in Fig.~\ref{fig:luminosityProfileChi2map_TXSB2353-003} show the
expected curves of n$=$4 Sersic profiles that match the central
1\arcsec\ of the light profile, with effective radii R$=$261~kpc
and R$=$15~kpc for TXS~2353-003 and NVSS~J21062-314003,
respectively. These effective radii are consistent with those we
found with two-dimensional fits using \verb+Galfit+ \citep[][]{Peng2002,
Peng2010}, which we ran on images that were slightly smoothed to
minimize pixel-to-pixel noise (by convolution with a 3$\times$3 pixel
Gaussian) and where we left all parameters free, except imposing that
n$=$4.

The  comparison between the expected profiles for n$=$4 and the
observations shown in
Fig.~\ref{fig:luminosityProfileChi2map_TXSB2353-003} illustrates that
only the inner regions of NVSS~J210626$-$314003 can be 
fitted with a de Vaucouleurs profile, with radius $R_e=15$~kpc. However, the
extended halo produces residuals greater than $3
\sigma$. For TXS~2353$-$003, the surface-brightness contrast between
the central regions and the extended outer halo is lower, resulting in
a very large size of $R_e = 261$~kpc for a de Vaucouleurs profile,
which matches the central and outermost isophotes, but also leads to
significant residuals at intermediate radii. This is in agreement with
the visual impression that only NVSS~J210626$-$314003 has a
distinctive, high-surface brightness core, whereas TXS~2353$-$003 is
more diffuse. High-resolution imaging with the Hubble Space
Telescope, for example,  would be needed for a full analysis of the
surface-brightness profiles in the inner regions of our galaxies. 
Nonetheless, these results already show that both galaxies show important
departures from simple de Vaucouleurs profiles at intermediate and large radii. 
Such extended wings are  characteristic of
central cluster galaxies  \citep[e.g.,][]{Schombert1987}.

\subsection{Line contamination}

Line emission in high-z radio galaxies can reach high equivalent
widths and might therefore be a significant contaminant in morphology
measurements through broadband filters \citep[][]{pentericci01,
  nesvadba08, targett11}. We will now investigate the significance of
this contamination in our analysis.
\begin{figure}
\includegraphics[width=0.24\textwidth]{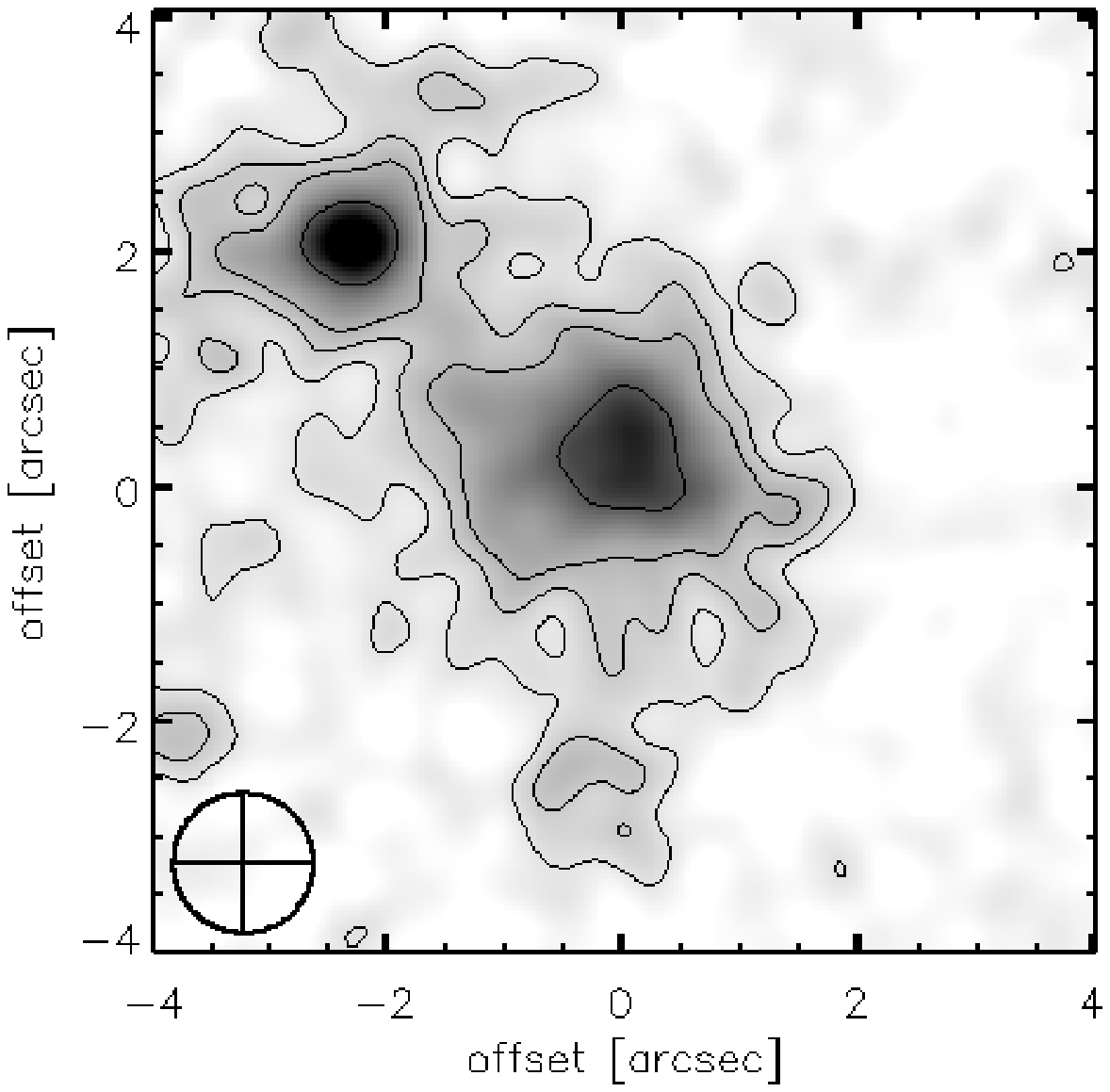}
\includegraphics[width=0.24\textwidth]{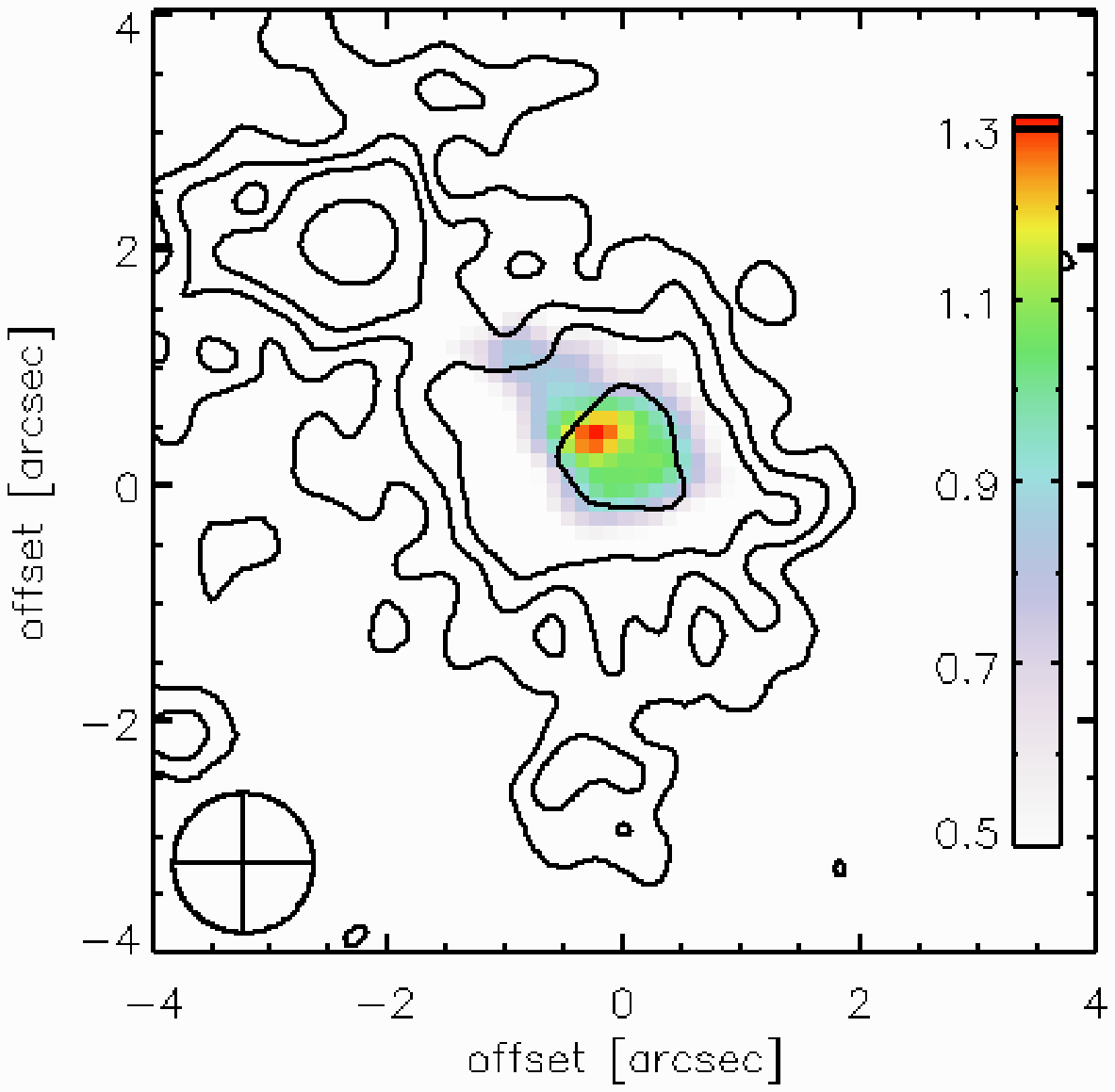}
\caption{{\it left:} ISAAC H-band broadband image of TXS~2353$-$003
convolved with a two-dimensional Gaussian to match the spatial
resolution of the H$\alpha+[NII]$ map shown in the right panel. {\it
right:} H$\alpha$ morphology of TXS~2353$-$003, convolved with a
two-dimensional Gaussian to emphasize the morphology of the faint
northeastern extension at a spatial resolution of 1.2\arcsec.
The color bar shows the contamination to the flux density measured in
the broadband filter from H$\alpha$ line emission in units of
$10^{-19}$ erg s$^{-1}$ cm$^{-2}$ \AA$^{-1}$ arcsec$^{-1}$. Contours show the
broadband K-band morphology and are identical to those shown in the
left panel.  Starting from the outermost contour, the broadband
fluxes represented by each contour are 1.1, 1.8, 2.5, and $5.4 \times
10^{-19}$ erg s$^{-1}$ cm$^{-2}$ \AA$^{-1}$ arcsec$^{-1}$ (3, 5, 7,
and 15$\sigma$ in the convolved image, where $\sigma=3.9\times
10^{-20}$ erg s$^{-1}$ cm$^{-2}$ \AA$^{-1}$ arcsec$^{-1}$). The
circles in the lower left of each panel show the size of the convolved
PSF. \label{fig:txs2353cont}}
\end{figure}

>From our SINFONI imaging spectroscopy we know the line fluxes of the
most prominent optical emission lines in both galaxies, H$\alpha$,
[NII]$\lambda\lambda$6548,6583, and
[OIII]$\lambda\lambda$4959,5007. In TXS~2353-003, the combined flux of
all emission lines in the H band reaches an average surface brightness
of $1.6\times10^{-15}$ erg s$^{-1}$ cm$^{-2}$ arcsec$^{-2}$ in a
2.15\arcsec\ aperture around the center of the galaxy, and $3\times
10^{-16}$ erg s$^{-1}$ cm$^{-2}$ arcsec$^{-2}$ along the northeastern
extension. Outside this region, line emission is not detected and must
therefore be below the 3$\sigma$ upper surface-brightness limit of
$1\times 10^{-16}$ erg s$^{-1}$ cm$^{-2}$ arcsec$^{-2}$.

The broadband morphology of TXS~2353$-$003 is shown in the left panel
of Fig.~\ref{fig:txs2353cont} as a gray scale image and contours. The
image was convolved with a two-dimensional Gaussian to a resolution of
1.2\arcsec, to match the size of the
seeing disk of the H$\alpha$ and [NII]$\lambda$6583 line image shown
in the right panel of Fig.~\ref{fig:txs2353cont}, and to enhance the
contrast of the faint extended emission in the broadband image
against the background noise. 

The right panel of Fig.~\ref{fig:txs2353cont} shows the
H$\alpha$+[NII] line image of TXS~2353$-$003, obtained by collapsing
the data cube along the spectral direction over wavelengths where
H$\alpha$+[NII] is detected. This makes the line morphology
measurement more robust than the maps obtained from Gaussian fits to
each individual pixel, in particular in the faint outer parts of the
emission-line regions, and allows the measurement of the upper limits
from the same image that is also used to estimate the emission-line
surface brightness.  Dividing by the 2700~\AA\ width of the H-band
filter of
ISAAC\footnote{http://www.eso.org/sci/facilities/paranal/decommissioned/
isaac/inst/isaac\_img.html}, we find that line emission contributes on
average 2.75$\times$ 10$^{-19}$ erg s$^{-1}$ cm$^{-2}$ arcsec$^{-2}$
\AA$^{-1}$ to the measured flux density in the H band near the center
of TXS~2353$-$003, and of $1.1\times 10^{-19}$ erg s$^{-1}$ cm$^{-2}$
arcsec$^{-2}$ \AA$^{-1}$ in the northeastern periphery. Outside this
extended emission-line region, the flux densities from emission lines
drop to below $3.7\times 10^{-20}$ erg s$^{-1}$ cm$^{-2}$
arcsec$^{-2}$ \AA$^{-1}$ (our 3$\sigma$ limit).  H-band surface
brightnesses at the location of the extended H$\alpha$ emission-line
region are above $2.5\times 10^{-19}$ erg s$^{-1}$ cm$^{-2}$
\AA$^{-1}$ arcsec$^{-2}$, which suggests that line contamination does
not dominate the overall broadband morphology of TXS~2353$-$003.

\begin{figure}
\includegraphics[width=0.24\textwidth]{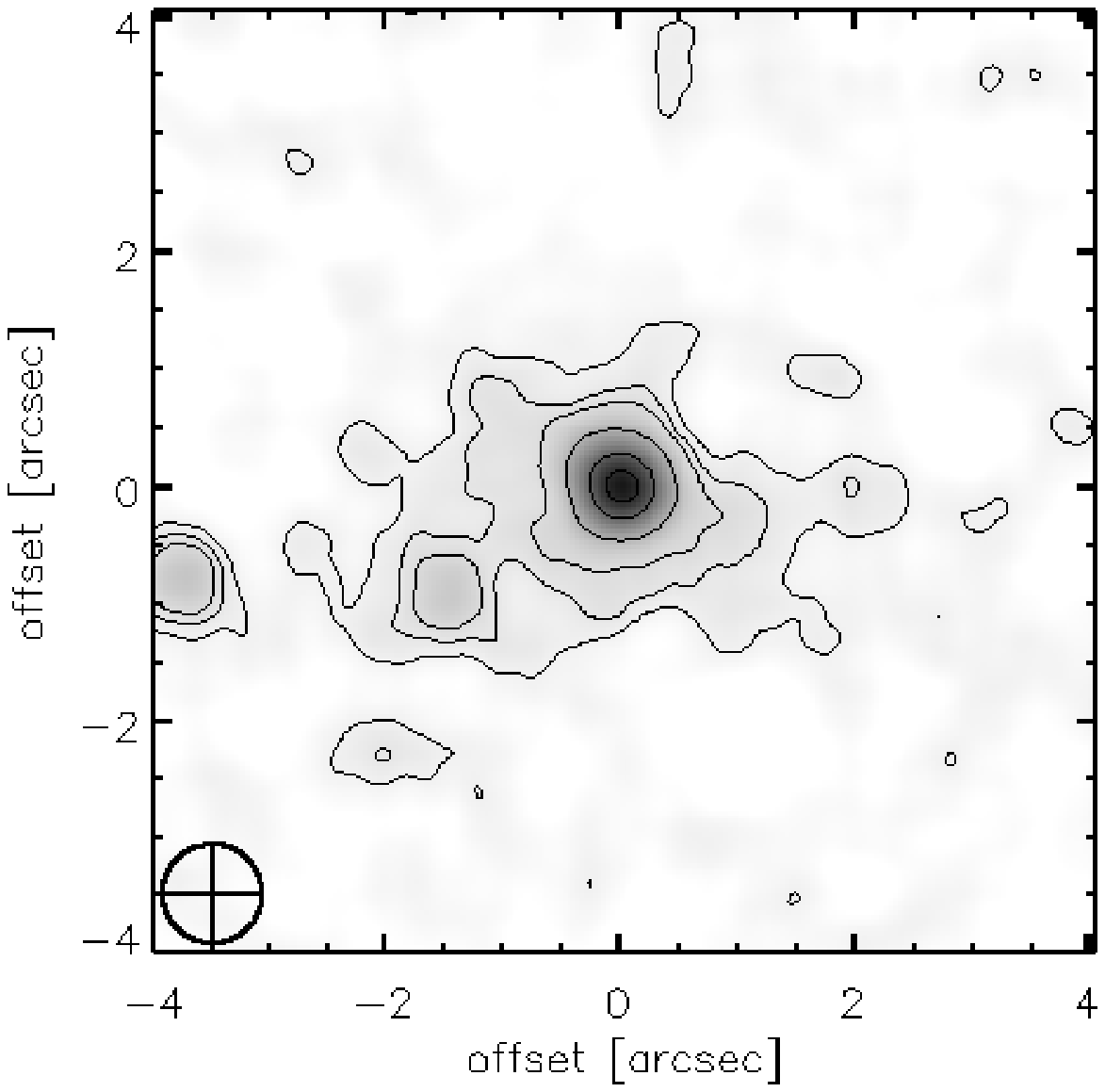}
\includegraphics[width=0.24\textwidth]{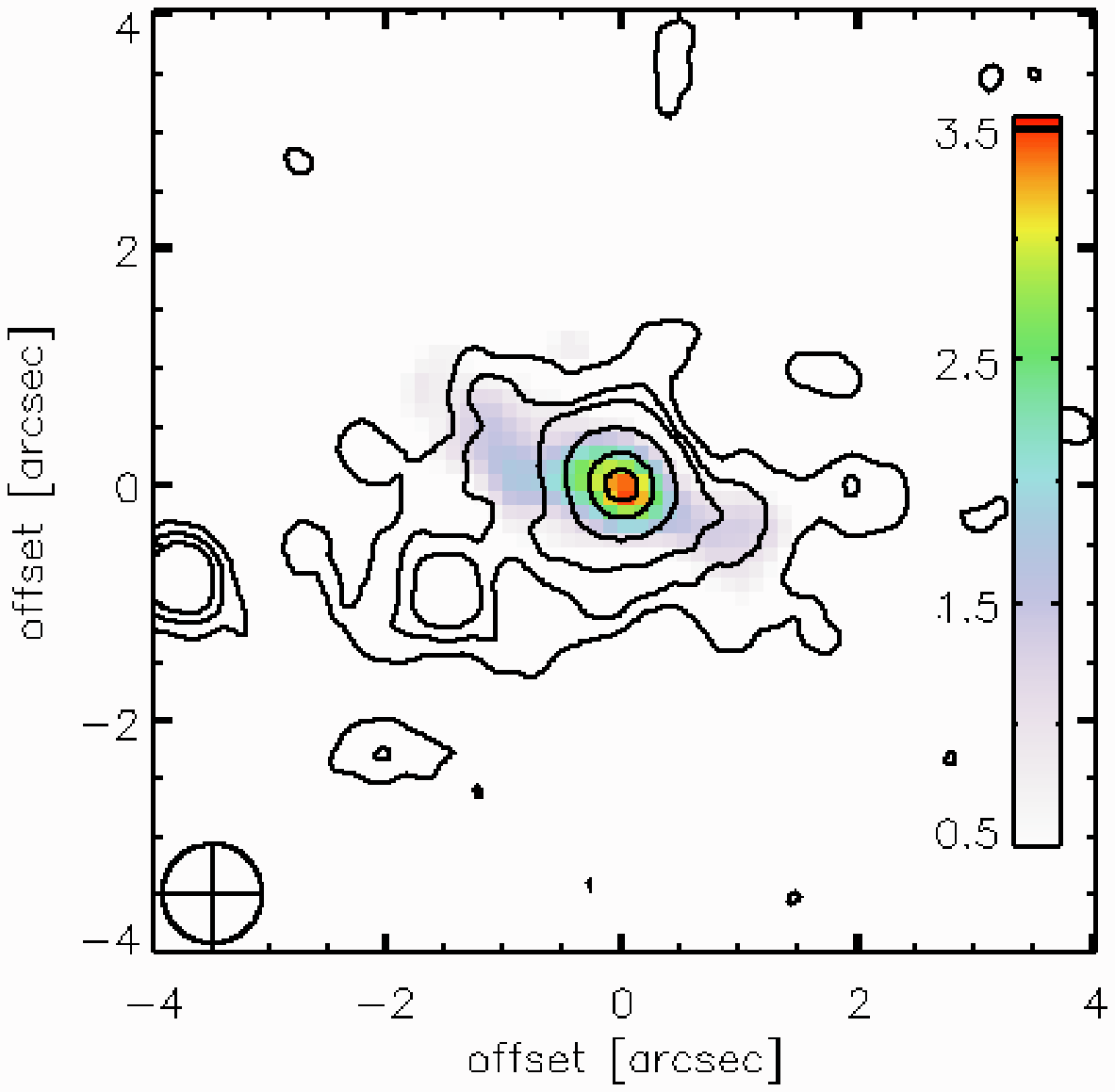}
\caption{{\it left} ISAAC K-band broadband image of
NVSS~J210626-314003 smoothed to the same spatial resolution as the
SINFONI imaging spectroscopy, 1\arcsec. {\it right} H$\alpha$
morphology of NVSS~J210626-314003. The color bar shows the
contamination to the flux density measured in the broadband filter
from H$\alpha$ line emission in units of $10^{-19}$ erg s$^{-1}$
cm$^{-2}$ \AA$^{-1}$ arcsec$^{-1}$. Contours show the broadband
K-band morphology and are the same as in the left panel. The
contour levels correspond to 3, 5, 7, 15, and 30$\times$ the root-mean
square of the broadband image smoothed to the SINFONI resolution of
1\arcsec, which has r.m.s.$=3\times 10^{-20}$ erg s$^{-1}$ cm$^{-2}$
\AA$^{-1}$ arcsec$^{-1}$. The circles in the lower left of each panel
show the size of the seeing disk.\label{fig:nvss2106cont}}
\end{figure}

The broadband image of NVSS~J210626$-$314004 is shown in
Fig.~\ref{fig:nvss2106cont} as a gray scale image with contours (left panel),
and as contours on top of the H$\alpha$ and [NII] line image (right
panel), which was constructed in the same way as described above for
TXS~2353$-$003. The line emission is closely approximated by
Gaussian line profiles, and we therefore use the emission-line map
derived from those fits and shown in
Fig.~\ref{fig:SINFONIobsNVSS2106}. In a 2.15\arcsec\ aperture around
the nucleus, we find an average surface brightness of H$\alpha$ and
[NII] of $6.9\times 10^{-16}$ erg s$^{-1}$ cm$^{-2}$ arsec$^{-2}$ and
$2.2\times 10^{-16}$ erg s$^{-1}$ cm$^{-2}$ arcsec$^{-2}$ in the
periphery, respectively. This corresponds to a surface flux density of
$2.3\times 10^{-19}$ and $9.4\times 10^{-20}$ erg s$^{-1}$ cm$^{-2}$
\AA$^{-1}$ arcsec$^{-2}$, respectively, after dividing by the filter
width of the ISAAC K-band filter, 3000~\AA$^{1}$. Outside the ridge of
extended line emission, the K-band flux densities from the lines drop
to below $3\times 10^{-20}$ erg s$^{-1}$ cm$^{-2}$ \AA$^{-1}$ arcsec$^{-2}$
(our 3$\sigma$ limit on the emission-line surface brightness).
 
Line emission in NVSS~J210626$-$314003 extends over regions of
broadband isophotes of $1.5\times 10^{-19}$ erg s$^{-1}$ cm$^{-2}$
and brighter (Fig.~\ref{fig:nvss2106cont}).  This comparison suggests
that, outside the brightest areas seen in the H$\alpha$ line image
near the galaxy center, line emission is not a major contaminant of
the broadband morphology in NVSS~J210626$-$314003. 

The diffuse Ly$\alpha$ halos discovered by \citet{villar02} and
\citet{villar03} would also produce fainter emission-line surface
brightnesses than we observe here. For a typical $SB_{Ly\alpha}=2-3
\times 10^{-17}$ erg s$^{-1}$ cm$^{-2}$ arcsec$^{-2}$, we would expect
H$\alpha$ surface brightnesses between $2-5\times 10^{-18}$ erg
s$^{-1}$ cm$^{-2}$ arcsec$^{-2}$, assuming typical
Ly$\alpha$/H$\alpha$ ratios of 8$-$13 \citep[][]{villar03}. Dividing
by the 2700\AA\ filter width of the ISAAC H-band filter, this would
correspond to $0.7-1.5\times 10^{-21}$ erg s$^{-1}$ cm$^{-2}$
\AA$^{-1}$ arcsec$^{-2}$, much lower than the surface brightness seen
in the broadband filter.

\section{H$\alpha$ candidates surrounding TXS~2353-003} 
\label{sec:environment}

Narrowband imaging is a standard way of identifying emission-line
galaxies at high redshifts, and has been successfully used to identify
over-densities of actively star-forming galaxies around high-redshift
galaxies for almost two decades \citep[e.g.,][]{chambers96,lefevre96,
Kurk2004a, venemans07, hatch11, koyama13, hayashi12, cooke14}. Most of
these observations were carried out for HzRGs at z$\gtrsim$2, where
H$\alpha$ falls in the K band, which is relatively free from telluric
night-sky lines.

We also obtained H$\alpha$ narrow-band imaging of TXS 2353-003, where
H$\alpha$ falls fortuitously into the NB 1.64 filter (although not
perfectly into the center). Our main goal was to search for
possible bright emission-line gas outside the small SINFONI field of
view of 8"x8", that would be associated with the environment rather
than the radio galaxy itself. This also enables us to identify emission-line galaxies that
are within the same dark-matter environment, and within a
  velocity range of $-$1580~km~s$^{-1}$ to $+$3000~km~s$^{-1}$ from
  the HzRG. Typical galaxy overdensities around HzRGs have velocity
  dispersions of $<$900 km s$^{-1}$ \citep[][]{venemans07}, which
  suggests that the asymmetry of the velocity range in our case does
  not hinder our detection of a potential overdensity of line emitters
  around the radio galaxy out to several times the velocity dispersion
  of the cluster. We present the results of both parts of this
narrow-line imaging project in this section.

\subsection{Identification of candidate H$\alpha$ emitters}
\label{ssec:hacandidates}
We identified candidate emission-line galaxies associated with the
dark-matter environment of TXS~2353-003 candidate  HAEs in two different ways. The first is
a method introduced by \citet{Bunker1995} that searches for a
flux excess in the narrow compared to the broadband image, and the second is the
direct inspection of the continuum-subtracted line image
described in \S\ref{ssec:narrowband}. 

For the flux excess method of \citet{Bunker1995}, we used SExtractor
v.2.5.0 \citep[][]{BertinArnouts1996} to construct catalogs from the
narrow- and broadband images, setting the parameters
\verb+DETECT_THRESH+ = 2.0 and \verb+ANALYSIS_THRESH+ = 2.0, while
leaving all others at their default values. We then take the positions
of the galaxies in the catalog extracted from the broadband image to
perform aperture photometry within 2\arcsec\ apertures in both
images. Using fixed apertures is appropriate for marginally or
unresolved galaxies like ours, and has the advantage of providing
errors that are independent of galaxy size \citep[see
also][]{Kurk2004a}. Selecting our candidates from the broadband image, rather
than the narrowband, is different from many other narrowband
searches, but makes our selection particularly robust, since it
requires independent detections in two images. The disadvantage is
that we might miss galaxies with particularly high emission-line
equivalent widths. However, comparison of the catalogs extracted from
the two images shows that this was not the case in our analysis.

We use the ``significance of excess'', $\Sigma$,
as defined by \citet{Bunker1995}, i.e., ``the number of standard
deviations between the counts measured in the broadband and the
number expected on the basis of the narrowband counts (assuming a
flat spectrum)'', to select the best H$\alpha$ candidates. In
addition, and again following \citet{Bunker1995} we require that the
sources fall above a given equivalent width threshold. Following
\citet{Kurk2004a}, we choose EW$_0\ge$50\AA\ and 25\AA. These thresholds
correspond to broad- and narrowband colors of 0.38 and 0.17, respectively. Three
sources with $\Sigma\ge3.0$  even have EW$_0\ge 100$\AA, including the
radio galaxy, which corresponds to a broad- to narrowband color of
0.69.

\begin{figure}
\begin{center}
\includegraphics[width=0.45\textwidth]{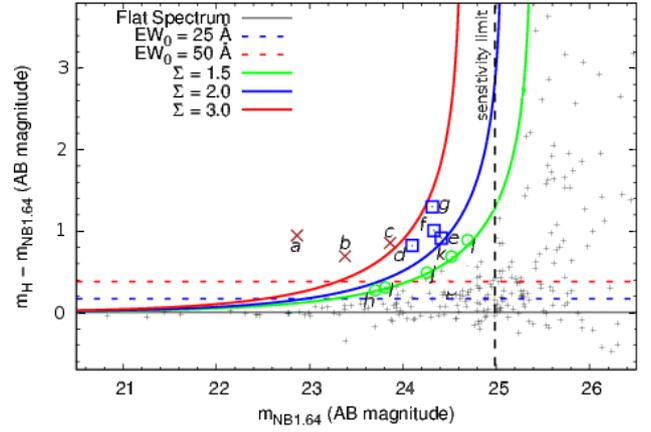}
\caption{Broad- to narrowband color-magnitude diagram of our candidate
  H$\alpha$ emitters. H$\alpha$ candidates are labeled with
  colored symbols, depending on the significance of their
  detection. Letters refer
  to the sources shown in Fig.~\ref{fig:thumbnailsHAE}. Source ``a'' is
  the HzRG.}

\label{fig:colorMagnitudeDiagram}
\end{center}
\end{figure}

\begin{figure}
\begin{center}
\includegraphics[width=0.48\textwidth]{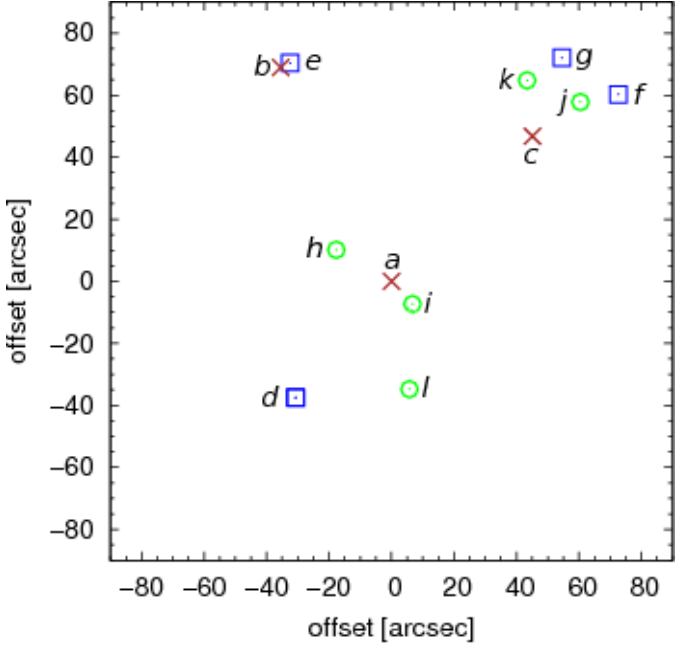}
\caption{Positions of the candidate H$\alpha$ emitters surrounding TXS~2353-003 on the sky. The HzRG is at the field center and
labeled with the symbol ``a''. Offsets are given relative to the
position of TXS~2353-003. North is up and east to the left.}
\label{fig:halphapos}
\end{center}
\end{figure}

The resulting color-magnitude diagram is shown in
Fig.~\ref{fig:colorMagnitudeDiagram}. We
find seven sources with $\Sigma\ge2.0$ and $EW_0\gtrsim 50$\AA. 
We also show another seven H$\alpha$ candidates
that have somewhat lower excess significances $\Sigma=$1.5-2.0. A
fraction of these are probably part of the overdensity of
TXS~2353-003. However, the fraction of misidentifications due to larger
spectral slopes, for example, is  likely to be larger among these sources,
which is why we restrict our analysis to the most robust candidates
above $\Sigma=$2.0.

\subsection{An overdensity of H$\alpha$ candidates around TXS~2353$-$003}
\label{ss:overdensityHaEmittersAroundTXSB2353-003}

Given the relatively good seeing of our narrowband data of 0.4\arcsec, most
H$\alpha$ candidates are spatially resolved
(Fig.~\ref{fig:thumbnailsHAE}) and we estimate H$\alpha$ equivalent
widths of 50$-$250~\AA\ in the rest frame, corresponding to H$\alpha$
fluxes between about 1 and a few $\times 10^{-16}$ erg s$^{-1}$
cm$^{-2}$. Their properties are listed in
Table~\ref{table:HalphaEmitters}. We did not correct the H$\alpha$
flux and equivalent widths for [NII] emission, which should fall into
the same filter. Observations of UV/optically 
selected galaxies at high redshift typically show very low
[NII]/H$\alpha$ line ratios on the order of  10\%
\citep[e.g.,][]{nmfs09,lehnert09,Queyrel2012}. Using
the conversion of \citet{Kennicutt1998} with a $1-100$~M$_{\odot}$
Salpeter initial mass function, we find star formation rates of 16-35
M$_{\odot}$ yr$^{-1}$ per galaxy, and a  moderate total
star formation rate of 138 M$_{\odot}$ yr$^{-1}$ in these galaxies
combined. These values were not corrected for extinction, so
  intrinsic star formation rates could be higher. For example,
  \citet[][]{garn10} and \citet{stott13}  typically find ~1 mag
  extinction in H-alpha meaning that intrinsic star formation rates
  are probably a factor of $\sim$2.5 times higher. The expected fraction
of low-redshift interlopers is much lower in our case than for
narrowband searches using Ly$\alpha$. The brightest emission lines
longward of H$\alpha$ are the Paschen and Brackett   hydrogen lines
in the near-infrared, with typically much lower equivalent widths than
H$\alpha$.

To quantify whether the observed number of H$\alpha$ candidates around
TXS~2353-003 may herald an overdensity of galaxies, we have to compare our data
with samples of H$\alpha$ emitters in the field at the same epoch.  A
blind redshift search for  HAEs at z~=~1.48 has recently
been performed by \citet{Sobral2013} through the NB~1.617 filter at
UKIRT, which has a very similar width of 210~\AA\ (compared to
250~\AA\ for our ISAAC filter), that we  use as reference for
the expected source density in the field. The HiZEL survey
\citep[][]{Sobral2009b, Sobral2010, Sobral2012, Sobral2013} gives an
estimate of the H$\alpha$ emitters that we can expect in a blank
field: \citet{Sobral2012} concentrated on H$\alpha$ emitters at
z~$\simeq$~1.47, very close to the z=1.49 of TXS~2353$-$003. They
found 295 NB emitters over the 0.67~deg$^2$ field of their main
analysis, and 411 over the entire 0.79~deg$^2$ field of their NB
observations. Among them, 190 objects also show an excess in deep NB
imaging centered on the [OII]$\lambda3727$ emission line and have
colors compatible with a photometric redshift z~$\simeq$~1.47. This
leads to densities of candidate HAEs in the
field of $\Sigma_{HAE} \simeq$~0.14~HAE~arcmin$^{-2}$ (with all NB
emitters) ; $\Sigma_{HAE} \simeq$~0.12~HAE~arcmin$^{-2}$ (with NB
emitters in their main field) and $\Sigma_{HAE}
\simeq$~0.08~HAE~arcmin$^{-2}$ (with NB emitters having an adequate
photometric redshift). These values include galaxies down to a
3$\sigma$ flux level of $7\times 10^{-17}$ erg s$^{-1}$ cm$^{-2}$,
compared to $1\times 10^{-17}$ erg s$^{-1}$ cm$^{-2}$ in our case.

Around TXS~2353$-$003, with an effective field of view of
5.4~arcmin$^2$ (neglecting the trimmed edges where our data did not
reach the full depth) we find six H$\alpha$ emitting candidates with
fluxes $> 7\times 10^{-17}$ erg s$^{-1}$ cm$^{-2}$ and significances
$\Sigma \ge$~2.0 in addition to the radio galaxy. This
 corresponds to a source density of 1.1~arcmin$^{-2}$, and an
 overdensity by a factor $\sim$8 relative to the field.

\citet{Kurk2004b} detected 28 HAEs in two ISAAC pointings around
  MRC~1138-262 at z~=~2.16 down to an H$\alpha$ flux limit of
  $2.5\times 10^{-17}$ erg s$^{-1}$ cm$^{-2}$, which corresponds to
  the same luminosity limit as the current analysis. They found a
  source density of 2.2~arcmin$^{-2}$.
Although our source density at z$=$1.5 is lower than that in the
z$=$2.16 overdensity around MRC~1138-262 in absolute terms, the
significance of the overdensity relative to the field at the same
epoch is high, because the star formation rate density at z~=~1.5 in
the field is already strongly declining for the highest H$\alpha$
luminosities \citep[e.g.,][]{Sobral2013}. 

Finding high surface-brightness gas out to the periphery of our
SINFONI data cube raises the question whether there is similarly
bright line emission outside the small,
8\arcsec$\times$8\arcsec\ field of view of SINFONI
(67~kpc$\times$67~kpc at z=1.49). We do not find any such emission
down to a 3$\sigma$ detection limit of $6\times 10^{-20}$ erg s$^{-1}$
cm$^{-2}$ \AA$^{-1}$ in the continuum subtracted narrowband
image. With the 250\AA\ width of the NB1.64 filter, and cosmological
surface-brightness dimming by a factor $(1+z)^4=39$, we are sensitive
to the equivalent of $\ge 60\times 10^{-17}$ erg s$^{-1}$ cm$^{-2}$
arcsec$^{-2}$ in nearby clusters. This is more than an order of
magnitude fainter than the surface brightness observed, e.g., in parts
of the filaments in the Perseus cluster  \citep[][]{Hatch2007}. 

\section{Comparison of TXS~2353-003 and NVSS~J210620-214003 with cluster central galaxies}
\label{sec:bcgs}

High-redshift radio galaxies do not often reside in solitude, but are surrounded by
overdensities of line or continuum emitters, suggesting that a
significant fraction of the HzRG population may be the progenitors of
the central galaxies of massive galaxy clusters
\citep[e.g.,][]{chambers96,lefevre96, Kurk2004a, venemans07,
galametz12}. \citet{wylezalek13} obtained Spitzer warm-mission IRAC
photometry at 3.6$\mu$m and 4.5$\mu$m of 200 HzRGs, including our two
sources. They find that, compared to the Spitzer UKDSS Ultra Deep
Survey \citep[][]{kim11}, 55\% of HzRGs are surrounded by $\ge2\sigma$
overdensities of galaxies with colors between the 3.6$\mu$m and
4.5$\mu$m channels that are consistent with redshifts z$\ge$1.3. Like
any method identifying galaxy overdensities, however, the IRAC color
selection only samples parts of the population of putative satellite
galaxies around HzRGs, and can therefore only provide  lower limits to
the actual density of such galaxies around a particular HzRG. It can
therefore only demonstrate the presence, but not the absence of such
an overdensity of galaxies around a given radio galaxy.

\citet{wylezalek13} show that TXS~2353-003 is surrounded by one of the
highest overdensities of IRAC-selected sources, with an excess of
$>5\sigma$, and consistent with the results of our narrowband search
presented in \S\ref{sec:environment}. NVSS~J210626-314003, however,
does not stand out as a source with a particularly dense environment
of IRAC-selected sources in the \citet{wylezalek13} sample, a 1.2
$\sigma$ excess.  Unfortunately, the observing program committee of ESO
only granted us   narrowband imaging of TXS~2353-003, not of
NVSS~J210626-314003, so that we were unable to do a narrowband search
around this galaxy. We note, however, that selecting a galaxy based on
its bright radio emission already leads to significant biases toward
galaxies in dense environments \citep[e.g.,][]{Best2000, ramos13,
hatch14}, and we argue in the following that the stellar component
of NVSS~J210626-314003 itself shows some of the characteristic
signatures of cluster central galaxies.

\subsection{K$-$z relationship}
\label{ssec:kmag}

\citet{kauffmann98} argued that the observed K-band magnitude can
be used to approximate the stellar mass of all but the most rapidly
growing galaxies out to redshifts z$\ge$2, with a scatter of about a
factor of~2. This is consistent with the empirical tight K-z
relationship between observed K-band magnitudes of powerful radio
galaxies and their redshifts \citep{lilly84,DeBreuck2002, Willott2003a,
  bryant09b} and the small mass range of these galaxies 
in the Spitzer photometric survey of HzRGs by
\citet{seymour07} and \citet{debreuck10}. 

\begin{figure}
\begin{center}
\includegraphics[width=0.45\textwidth]{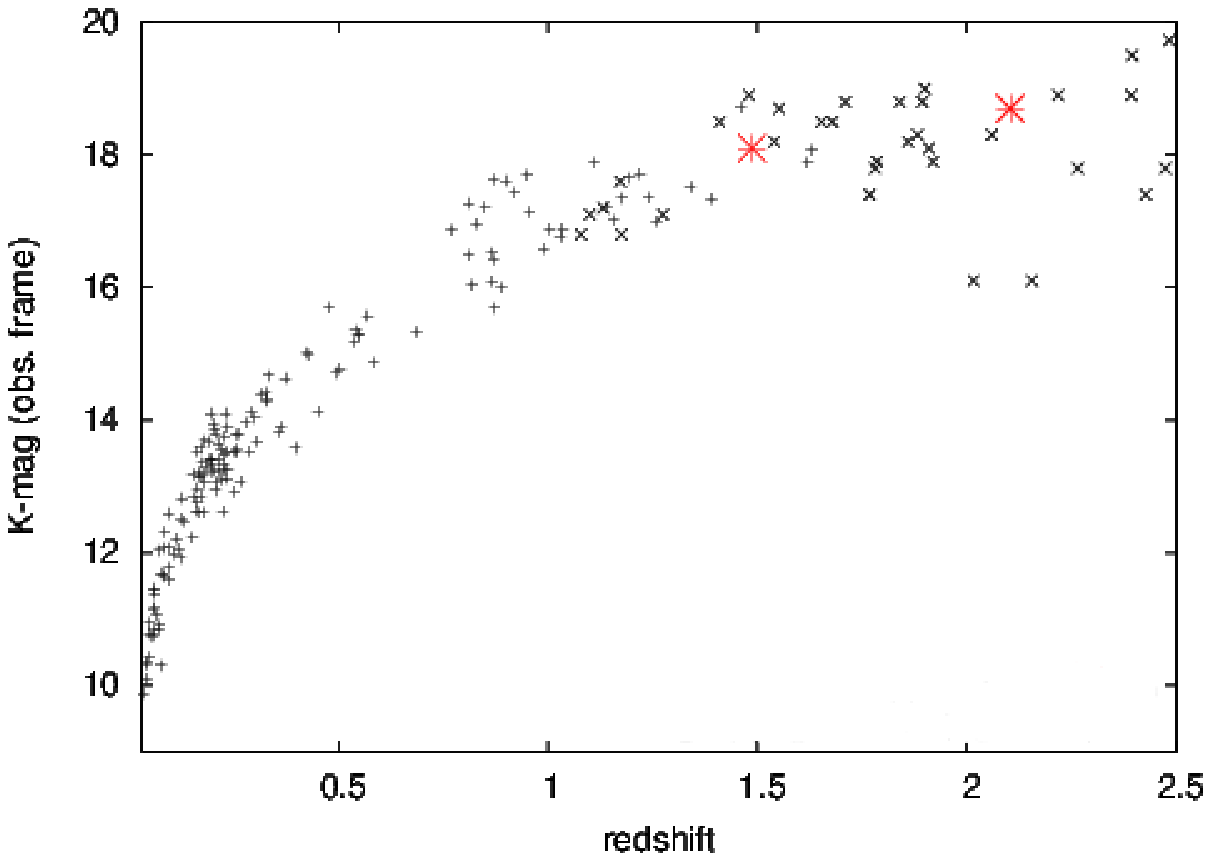}
\caption{Observed $K-$band magnitude of brightest-cluster galaxies
(shown as `+') and radio galaxies (shown as `x') as a function of
redshift \citep[plot reproduced from][and slightly extrapolated to
z=2.5]{Lidman2012}. Some of the data points were taken from
\citet{stott08} and \citet{stott10}. The large red stars illustrate
that TXS~2353$-$003 and NVSS~J210626$-$314003 fall well within the
expected range of $K-$band magnitude, and also have typical $K-$band
magnitudes for HzRGs at their redshifts \citep[black crosses][see also
\citealt{DeBreuck2002, Willott2003a, bryant09b}]{seymour07}. At
z$\ge2$, H$\alpha$ and [NII]$\lambda\lambda$6548,6583 fall into the
K-band filter and can affect the continuum magnitudes by up to about
0.3~dex in extreme cases \citep[][]{nesvadba06}.}
\label{fig:K-redshift}
\end{center}
\end{figure}

In Fig.~\ref{fig:K-redshift} we show where TXS~2353$-$003 and
NVSS~J210626$-$314003 fall relative to the K-z relationship of
\citet{seymour07} and \citet{debreuck10}, and that of
\citet{Lidman2012}, who found a similar relationship for brightest cluster galaxies (BCGs). This
figure shows the good agreement of our targets with both samples, as
well as the self-consistency of both samples with each other (as
expected if HzRGs are the progenitors of BCGs).

Encouraged by the results of \citet{kauffmann98, seymour07}, and
\citet{debreuck10}, which, when considered together, suggest that the
observed K-band magnitude of HzRGs should scale with the stellar mass,
we used the K-band magnitudes of $z\sim2$ HzRGs in the
\citet{seymour07} sample and the stellar mass estimates given in the
same paper, to obtain a simple empirical relationship between stellar
mass, $M_{stellar}$, and observed K-band magnitude, $m_K$,
$\log{(M_{stellar}/M_{\odot})}= 6.18-0.189\ m_K$. We find a 1 $\sigma$
scatter of about 0.3~dex in mass, consistent with that previously
found by \citet{kauffmann98}. We note that we are using observed
magnitudes from galaxies in a similar redshift range, and therefore
do not need a k-correction.

This allows us to derive a rough mass
estimate for our two sources. For TXS~2353$-$003 and
NVSS~J210626$-$314003, K-band magnitudes of 18.1$\pm$0.3 mag and
18.7$\pm$0.2 mag suggest stellar masses in the range
$\log{(M^{2353}/M_{\odot})}=11.1-11.4$ and
$\log{(M^{2106}/M_{\odot})}=11.2-11.5$, respectively. While such an
approach should not replace more detailed analyses of the continuum
emission where additional photometric constraints are available, it
does highlight that our two galaxies fall into the typical range of
stellar masses of powerful radio galaxies at these redshifts.

\subsection{Surface-brightness profiles}

In Section~\ref{ssec:results:continuum} we showed that both galaxies
are spatially resolved in our ground-based imaging, and that both have
surface-brightness profiles that are not consistent with the pure
Sersic profiles of early-type galaxies (n$=$4). The central arcsec of
the profile of NVSS~J210626$-$314003 can be fitted with a de
Vaucouleurs profile with R=15~kpc, but fainter extended emission
persists at larger radii. TXS~2353-003 can only be fitted with a de
Vaucouleurs profile when assuming an unphysically large radius of
261~kpc.

\citet{targett11} analyzed the K-band morphologies of 13 radio
galaxies at z$=$1.5$-$2.5 observed with UKIRT during very good seeing,
and found that most sources are well fit with Sersic indices n$=3-5$
and effective radii between 5 and 10~kpc. This is in contrast to the
findings of \citet{pentericci01} with HST/NICMOS through the F160W
filter that only 5 out of 19 sources have well-established de Vaucouleurs
profiles \citep[although the][sample does extend to higher redshifts
than that of \citealt{targett11} and, with H-band observations,
samples much shorter wavelengths]{pentericci01}. Only one source of
the \citet{targett11} sample, 0128$-$264, has an effective radius
R=15~kpc, as does  NVSS~J210626$-$314003. This galaxy is
also included in our parent sample with SINFONI data. It has a small
emission-line region, although elongated along the radio jet axis, and
a very extended (294~kpc) radio size, resembling the two sources
discussed here (Nesvadba et al. 2015, in prep.).
 
Massive early-type galaxies at high redshift appear to be generally
characterized by their greater compactness compared to similar, more nearby
 galaxies \citep[][who found $R_e \le 1$~kpc and 0.9~kpc,
  respectively ]{Daddi2005, vanDokkum2008}. Typically, these galaxies
are only resolved with deep, high-resolution HST imaging, showing no
deviations from S\'ersic laws \citep[e.g.,][]{Szomoru2010,
  Cassata2010}. The HzRGs of \citet{targett11}, which have relatively
large radii (average 8~kpc) compared to submillimeter galaxies
(average 3~kpc) at similar redshifts, show an average decrease in size
by a factor 1.5 compared to equally massive galaxies at low
redshift. A size decrease by a factor of  $\sim2$ has also been found by
\citet{pentericci01} for their HzRGs that are well modeled with a de
Vaucouleurs profile.

Extended stellar envelopes are a primary characteristic of central
cluster galaxies in the low-redshift Universe \citep[cD galaxies;
e.g.,][]{Schombert1987}, and likely formed from the debris of repeated
accretion of several satellite galaxies. This interpretation also
agrees with the finding of \citet{targett11} that HzRGs fall onto the
same Kormendy relationship between central surface brightness and
effective radius as massive low-redshift galaxies, which, as stated by
the authors, would imply that the size increase from z$\sim$2 to today
must be accommodated by an increase in stellar mass at large
galactocentric radii, as could be produced by repeated accretion of
satellites. \citet{best98} identified similar structures in a stack of
12 radio galaxies from the 3CRR sample at z$=0.6-1.4$, but to our
knowledge, the only example of a HzRG with extended stellar halo at
greater redshift currently known in the literature is MRC~1138$-$262
at z=2.2 \citep[the Spiderweb Galaxy,][]{miley06}.  \citet{Hatch2008,
hatch09} discussed the envelope around this source in detail, finding
that the rest-frame UV colors of the faint continuum emission are
consistent with a stellar envelope (rather than scattering from a
dusty gaseous halo) with a formation history that is 
coeval with that of the high-surface brightness central regions. This
is in contrast to semi-analytical simulations which postulate that
such envelopes assembled gradually through dry mergers over
cosmological timescales \citep[e.g.,][]{delucia07}.

\section{The nature of the extended ionized gas}
\label{sec:nature}

A good alignment between jet and gas is  typical of distant radio
galaxies \citep[the alignment effect; e.g.,][]{cimatti93}, and this
alignment was also among the prime arguments of \citet{collet14} and
of \citet{Nesvadba2006,nesvadba08} that jet and gas are physically
related in their SINFONI samples of HzRG.  They also found very large,
super-gravitational velocity gradients reaching to more than~1000~km
s$^{-1}$ associated with the most powerful radio sources, and broad
line widths in the extended gas with FWHM $\gtrsim$ 800 km
s$^{-1}$. They considered this as evidence of fast, AGN-driven
outflows in HzRGs.

None of this applies here. Velocity gradients are well below 1000 km
s$^{-1}$, line widths outside of the central regions are only on the order of 
200-300~km s$^{-1}$, and the mismatch in position angle between jet
and gas is striking. This mismatch makes it difficult to postulate
that jet cocoons are driving the gas out as they expand through the
ambient gas, because it appears unavoidable in this model that
cocoon (and hence, jet) and gas are cospatial, both in three
dimensions and in projection.

 The presence of this off-axis component of line emission 
is worth discussing, as is  the absence of bright, extended emission-line
regions along the jet axis. TXS~2353$-$003 may have a faint, not well
resolved emission-line component along the jet axis and within the
galaxy itself. However, this does not stand out compared to the gas in
other regions of the galaxy, and is therefore difficult to associate
uniquely with the radio jet. We find no similar emission at larger
radii, including the cluster scales that we cover with ISAAC around
TXS~2353-003 (\S\ref{sec:environment}).

The absence of bright gas on scales much larger than the host galaxy
cannot be a mere observational effect, since the sensitivities of these
data are comparable to the overall sample, and the surface brightness
of the line emission in our targets is not significantly lower than in
other HzRGs with SINFONI data. However, for the largest sources with
'regular' line emission, in particular MRC~2104-242 at z=2.5, the
extended line emission can become very faint and is concentrated along
long, thin, very filamentary structures, suggesting that the warm gas
is either being heated to temperatures T$\gg 10^4$ K, or becoming
strongly dispersed as the cocoon continues to expand (Nesvadba et
al. 2015, in prep.; see also \citealt{pentericci01} who observed the
same gas as faint, extended filaments in their HST/NICMOS broadband
imaging).

We stress that finding extended gas around HzRGs that is unrelated to
the radio jets is by itself not uncommon. Many HzRGs are surrounded by
diffuse halos of warm ionized gas \citep{villar03}, however,
dense, high-surface-brightness Ly$\alpha$ emission consistent with the
H$\alpha$ surface brightnesses that we observe here is only found
within the jet cocoon in the galaxies of \citet{villar03}
H$\alpha$ associated with the diffuse ionized gas would be about 2
orders of magnitude fainter than what we observe.

We will now discuss several hypotheses, from AGN orientation to galaxy
interactions to extended gas disks or filaments within the radio
galaxy itself or its immediate surroundings, to explain the nature of
this  gas.

\subsection{Partial AGN illumination of ambient halo clouds}

The line ratios of both sources suggest that they are photoionized by the
central AGN (\S\ref{ssec:results:linemission}). This could imply that
they do not represent a distinct structure, but are merely part of a
population of ambient clouds that are distributed over large solid
angles, and of which a subset is being lit up as it  intercepts
the quasar ionization cone. Filaments of neutral gas have also been
observed around some HzRGs \citep[][]{vanOjik1997, jarvis03, wilman04,
 debreuck05, nesvadba09, emonts14}.  

However, not all HzRGs show signatures of neutral gas clouds in their
halos.  \citet{vanOjik1997} found Ly$\alpha$ absorption only
in galaxies with radio sizes $<$50~kpc, in stark contrast to the large
radio sizes, $\ga 200$~kpc, of TXS~2353-003 and NVSS~210626-314003.
\citet{binette00} showed from an analysis of the ionization properties
that these absobers must also lie at larger radii than the radio
source. In the more compact radio galaxies, Ly$\alpha$ absorbers are
found not only  against the nucleus, but also against more extended
Ly$\alpha$ emission around the galaxy \citep[][]{Kurk2004a}, which
suggests that this result holds for large solid angles around the
radio galaxies, and not just along the radio jet axis.

This scenario would also require a misalignment between radio jet axis
and quasar illumination cone, as is possible because of jet
precession, or a misalignment between jet axis and the normal of the
torus, for example. \citet{drouart14} recently argued, based on Spitzer
mid-infrared imaging and the core-to-lobe fraction of the radio
sources, that the unified model must hold for the global population of
HzRGs, but the presence of nuclear broad H$\alpha$ emission in a few
sources does also suggest that this might not be the case for each
individual HzRG \citep[][]{nesvadba11a}.

Alternatively, we may be seeing gas in an old cocoon from a previous
radio-loud activity phase of the super-massive black hole. Jet
precession has been invoked to explain the  X-shaped radio
sources at low redshift, where about 5-10\% of FRII radio sources have
two pairs of radio lobes \citep[][]{leahy02}. The second bar of the
``X'', typically secondary pairs of fairly diffuse, low
surface-brightness wings in centimeter radio continuum imaging, may be the
relic of a previous burst of nuclear activity along a different jet
axis. Detecting such low surface-brightness emission directly would be
very challenging at high redshift.

This scenario, however, also has its
difficulties. \citet[][]{kaiser02} argued that clouds in relic cocoons
will be destroyed by shocks within a few $10^6$ yrs, which is of the
same order as the ages of bright radio sources in high-redshift
galaxies \citep[e.g.,][]{blundell99}. In the cocoon model, dense
clouds are confined by the high pressure of the cocoon material, and
should disperse within a sound-crossing time once the jet has stopped
maintaining the cocoon pressure high \citep[e.g.,][]{fabian87}. From
the size of the two lobes in our sources, 310~kpc and 190~kpc, and
assuming a jet advance speed on the order of $0.1c$, we can roughly
constrain that the feeding of a putative older jet component must have
ended at least about $1\times 10^7$ yrs ago in our two sources, so
that most of the emission-line clouds should have already been
destroyed or evaporated. The relatively old age of our two sources
does not make them good candidates for seeing relic cocoons, in
particular since many younger radio sources, which should in principle
have brighter relic cocoons, do not show evidence of two sets of radio
lobes or extended emission-line regions.

We conclude that neither partial AGN illumination of a general
population of ambient clouds  nor repeated cycles of radio-loud AGN
activity with different jet orientations appears to be a good
explanation for the nature of our sources.

\subsection{External gas supply from a satellite galaxy?}


Gas transfer from a satellite galaxy undergoing accretion onto the
HzRG, as previously observed in the z=3.8 HzRG 4C60.07
\citep[][]{Ivison2008}, is another interesting hypothesis for the
origin of this misaligned gas in our two sources. In this case, the
extended line emission could trace a gas tail produced by ram-pressure
stripping or tidal forces. However, only TXS~2353$-$003 has a
companion along the direction of the extended gas, but at a larger
distance from the radio galaxy. NVSS~J210626-314003 has no obvious
companion within several tens of kpc that can be associated with the
extended emission-line region.

It is uncertain whether the galaxy near TXS~2353-003 (Fig.~1) is
physically associated with the radio galaxy, although its infrared
colors as measured with IRAC are consistent with a redshift z$\ga$1.3
using the criteria of \citet{papovich08}
(\S\ref{sec:environment}). However, we detect no H$\alpha$ line
emission down to $3.3 \times 10^{-18}$ erg s$^{-1}$ cm$^{-2}$. Tidal
forces during galaxy interactions  produce extended tails, and
also funnel gas toward the nuclei of the galaxies
\citep[e.g.,][]{Barnes1996}, fueling intense nuclear
starbursts. Likewise ram pressure should not only strip parts of the
ISM, but also compress the gas along the head of the infalling galaxy,
enhancing the gas surface brightness within the galaxy, and
potentially star formation \citep{Kapferer2008}.

It appears therefore difficult for a galaxy to produce a tail of
$\sim 10^9$ M$_{\odot}$ of ionized gas without sustaining significant
star formation.  Our upper limit of H$\alpha$ flux corresponds to
star formation rates on the order of 10 M$_{\odot}$ yr$^{-1}$ \citep[using
  the][calibration and neglecting extinction]{Kennicutt1998}. Since
the source is bright in the continuum at similar wavelengths, the
H$\alpha$ equivalent width must also be low.  Although gas, dust, and
young stars are not necessarily co-spatial
\citep[][]{calzetti97}, this would imply here that the galaxy, in
spite of losing considerable fractions of its ISM in the putative
tail, is also maintaining a highly efficient obscuring dust screen
around its star-forming regions. This appears contradictory, in
particular for high-z galaxies, which typically have extended star
formation in multiple knots across the galaxy
\citep[e.g.,][]{nmfs09}. Other HzRGs, e.g., MRC~1138-262 at z~=~2.16,
are already known to be surrounded by companions within a few tens  of
kpc that  are already on the red sequence \citep[][]{Kodama2007} and
lack line emission \citep[][]{Nesvadba2006,Kuiper2011}. It
thus appears unlikely that external gas supply from a satellite galaxy
is a good explanation of the extended emission-line regions in our
galaxies.

\subsection{Extended gas disks within the radio galaxy}

\citet{Best2000} pointed out that radio galaxies with very extended
radio sources are particularly good candidates for being cluster
central galaxies because the pressure from the Mpc-scale intracluster
medium  boosts the luminosity of the radio source even for
comparably extended jets \citep[see also][]{athreya98, klamer06}. Many
authors have previously suggested that the progenitors of brightest
cluster galaxies are likely to be found among HzRGs
\citep[e.g.,][]{pentericci01,miley06,hatch09}, and \citet{wylezalek13}
 find from IRAC imaging that TXS~2353-003 is even associated
 with the most strongly pronounced overdensity within their
 sample.

The gas that we  see here may actually have much in common with the
extended gas disks or filaments that are  found in and near
cluster central galaxies in the more nearby  Universe.
Mismatches between the position angles of jet and gas are 
fairly common in  about 30\% \citep{McDonald2010} of cool-core
galaxy clusters with extended H$\alpha$ filaments. A particularly
clear example, where the warm ionized gas appears to avoid the
cavities that the radio jet has inflated in the intracluster medium is
Abell 1795, in particular at the highest emission-line surface
brightnesses \citep{vanBreugel1984}. In total, such structures have up
to a few $10^{10}$~M$_{\odot}$ \citep[e.g.,][]{Salome2006} of warm and
cold gas with T$\le 10^4$~K, typically dominated by  cold molecular
gas. We have already argued in \S\ref{sec:bcgs} that our sources
follow the overall trends found in BCGs between H$\alpha$ luminosity and surface
luminosity, size, and ratios of low-ionization emission lines.

Imaging spectroscopy of cool-core clusters with bright optical line
emission \citep[e.g.,][]{Hatch2006,Wilman2006,Hatch2007,Wilman2009,
  Farage2010, Farage2012} shows that the gas has typically FWHM $\sim
100-300$~km s$^{-1}$ outside the near-nuclear regions, with a
characteristic broadening toward the center, with typical FWHM $\ge$
500~km s$^{-1}$ and up to the values we find near the center of
NVSS~J210626$-$314003 and TXS~2353$-$003
(\S\ref{ssec:results:linemission} and
Figs. \ref{fig:SINFONIobsTXS2355} and \ref{fig:SINFONIobsNVSS2106}). However,
regular velocity gradients of a few 100~km s$^{-1}$, similar to what we
observe in our sources, are found only in some BCGs, e.g., Abell~262
\citep{Hatch2007} or 2A~0335$+$096 \citep{Farage2012}. Alternatively,
disks with well-ordered velocity fields, perhaps more like in
NVSS~J210626-3140003 than in TXS~2353$-$003, have also been found, for
example in Hydra~A \citep{Hamer2014}. Such disks are also typically
not aligned with the radio jet axis, but can be illuminated by the
central AGN, in particular if the opening angle of the AGN
illumination cone is large. 

The observed velocity gradients in our two sources are in fact
consistent with rotational motion in a gravitational potential as
suggested by the stellar mass estimates of $1.5-3\times 10^{11}$
M$_{\odot}$ (\S\ref{ssec:kmag}). A dynamical mass estimate, $v$, can
be derived by setting $v = \sqrt{M \sin{i}^2 G / R}$, where $M$ is the
stellar mass, $R$ the radius of the putative disk, $i$ the inclination
angle, and $G$ the gravitational constant. With R$=$10.5~kpc, measured
for the very regular velocity field of NVSS~J210626-314003, we find
v$=250-350\ \sin{i}^{-1}$ km s$^{-1}$, compared to 220 km s$^{-1}$
observed. Even if the gradients
were dominated by radial outflow motion, the small velocity range implies that most of this gas is
unlikely to escape.

This is consistent with the finding that the gas extends over similar
radii to the faint continuum wings in the ISAAC image of TXS~2353-003
on both sides, and in NVSS~J210626-314003 on the southwestern side.  If the
wings in the continuum surface brightness profiles of our two sources
represent extended stellar halos \citep[][]{hatch09}, perhaps
originating from a phase of rapid accretion of many satellite galaxies
in the very early evolution of the HzRGs \citep[][]{burkert08}, then
we may be seeing leftover gas that is settling down after this
phase. \citet{McDonald2010} suggested that the condensation of such
gas into fairly dense clouds could be accelerated by weak shocks
caused by the passage of a companion galaxy, as may be the case for
the putative companion in TXS2353-003.

Weak shocks could also be produced by the vestiges of the relic cocoon
that initially formed when the jet was passing through the inner
regions of the halo. This scenario has some resemblance to the
cyclical AGN feedback scenarios invoked, e.g., by
\citet{PizzolatoSoker2005} or \citet{AntonuccioDeloguSilk2010} to
explain extended gas disks and filaments in nearby cluster central
galaxies. \citet{nesvadba10,nesvadba11b} argued that isolated
  radio galaxies with extended gas disks may also require repeated,
  episodic AGN activity, broadly akin to the \citet{PizzolatoSoker2005}
  model, in order to understand the observed properties of the gas.
Winds associated with the large radio jets we observe may have
dispersed the ISM of these galaxies just a few $10^7$~yrs ago, and
parts of this gas may now be raining back onto the galaxy, possibly
feeding a new feedback cycle. The age of the Universe at z=1.5-2
  is already $>3$~Gyr, so that repeated activity cycles are possible
  for duty cycles of a few times $10^8$ yrs, as seems appropriate at low
  redshift \citep[][]{PizzolatoSoker2005}. The gas cooling times
could be lowered by compression of the outflow and diffuse ambient gas
that is being swept up as the two bubbles of the AGN cocoon continue
to inflate in the form of a momentum-driven wind after the feeding from
the AGN has ceased \citep[][]{kaiser02}. This cyclical feedback may
contribute to forming the hot intracluster X-ray gas in massive galaxy
clusters today, which presumably took place around redshift z$\ge$2
\citep[][]{nath02, mccarthy08}. Energy injection through repeated
episodic AGN activity has also been invoked as an explanation of how
the X-ray halos surrounding individual massive early-type galaxies can
be maintained over a Hubble time \citep[e.g.,][]{mathews03,best06}.

\section{Summary}
\label{sec:summary}
We have presented an analysis of rest-frame optical imaging spectroscopy
and deep broadband near-infrared imaging of two radio galaxies at
z$\sim$2, which have extended emission-line regions that are strongly
misaligned relative to the axis of the radio jets. This is in stark
contrast to the majority of powerful radio galaxies at similar
redshifts, where the gas is aligned within 20-30$^\circ$ of the jet
axis, as expected from a cocoon of turbulent gas that is being
inflated by the expanding radio jets. For one source, TXS~2353-003 at
z$=$1.5, we also present H$\alpha$ narrowband imaging through the
[FeII]1.64 filter of ISAAC, finding an overdensity of H$\alpha$
candidates which resembles those of radio galaxies at higher redshift. 

The gas in both galaxies is less perturbed than in radio galaxies with
more typical gas properties. In particular, line widths in the extended
gas are lower and, with FWHM $\sim$200-300 km s$^{-1}$ comparable to
those in the extended gas disks and filaments surrounding brightest
cluster galaxies at lower redshifts. Overall, we find remarkable
similarities to the BCGs in low-redshift cool-core clusters, ranging
from the extended, faint continuum halos in both galaxies that are
atypical for massive high-redshift galaxies except for the particularly
well-studied Spiderweb Galaxy MRC~1138-262, where such structures
are interpreted as extended stellar halos broadly akin to cD galaxies
at lower redshifts. Both galaxies fall onto the relationship between
K-band magnitude and redshift for brightest cluster galaxies and
HzRGs, suggesting they are among the most massive galaxies at their
epoch. We note that, given that HzRGs are now generally
  considered to be the progenitors of BCGs, the distinction between
  both classes is not mutually exclusive, and our two sources may
  simply be further evolved than many other HzRGs, but still following the same
  overall evolutionary path.

We discuss several scenarios for the nature of this gas, finding that
simple illumination effects in otherwise typical HzRG halos is not a
good explanation, and neither are galaxy interactions. Generally
speaking, our results support the hypothesis that the extended
line emission arises from extended gas disks or filaments within or
near the radio galaxy, by analogy with broadly similar structures in
nearby cluster central galaxies. This could be evidence for cyclical
AGN feedback, which has been discussed on several occasions for nearby
clusters.

\section*{Acknowledgments}

We are very grateful to the staff at Paranal for having carried out
the observations on which our analysis is based. We also thank the
anonymous referee for the detailed comments which helped improve the
paper. CC wishes to acknowledge support from the Ecole Doctorale
Astronomie \& Astrophysique de l'Ile de France. Parts of this research
were conducted by the Australian Research Council Centre of Excellence
for All-sky Astrophysics (CAASTRO), through project number
CE110001020.

\bibliographystyle{aa}
\bibliography{bibliography,hzrg}

\onecolumn

\begin{table}
\begin{center}
\caption{Properties of the two galaxies discussed in this study.}
\begin{tabular}{lcc}
\hline 
                                      & TXS~2353$-$003                          & NVSS~J210626$-$314003 \\
\hline 
Redshift                              & 1.487                                  & 2.104 \\
D$_L$ [Gpc]                           & 10.8                                   & 16.6 \\
radio power [$10^{28}$ W Hz$^{-1}$]    & 3.7                                    & 4.2\\
radio size [arcsec]                   & 38.8                                   & 24.2  \\
radio size [kpc]                      & 328                                    & 201 \\
K-band magnitude [mag]                & 18.1$\pm$0.3                           & 18.7$\pm$0.2\\
PA offset between jet and gas         & $\sim 90^\circ$                          & $\sim 60^\circ$ \\
F([OIII]$\lambda$5007)\tablefootmark{a}         & \dots                                   & $66 \pm 13$ \\
F(H$\beta$)\tablefootmark{a}                    & \dots                                   & $5.2 \pm 1$   \\
F([NII]$\lambda$6548)\tablefootmark{a}          & $2.8 \pm 0.6$                           & $1.2 \pm 0.2$ \\
F([NII]$\lambda$6583)\tablefootmark{a}          & $8.3 \pm 1.7$                           & $3.7 \pm 0.7$ \\
F([SII]$\lambda$6716)\tablefootmark{a}\tablefootmark{b}        & $1.4 \pm 0.5$                           & $2.3 \pm 0.5$ \\
F([SII]$\lambda$6731)\tablefootmark{a}          & $4.1 \pm 1.0$                           & $2.2 \pm 0.5$ \\
F(H$\alpha$)\tablefootmark{a}                   & $12.0 \pm 2.4$                          & $14.7 \pm 3.0$ \\
${\cal L}$(H$\alpha$) $[10^{43}$ erg s$^{-1}$] & 1.7  & 4.8 \\
ionized gas mass [$10^8$  M$_{\odot}$]  & 2.4  & 9.3 \\

\hline 
SINFONI rms\tablefootmark{c}   & 4.3 & 6.2 \\
\hline 
\end{tabular}
\label{tab:propertiesUnalignedGalaxies}
\end{center}
\tablefoottext{a}{All fluxes are given in units of $10^{-16}$ erg s$^{-1}$ cm$^{-2}$.}
\tablefoottext{b}{[SII]$\lambda6716$ in TXS~2353$-$003 is affected by a telluric line.} 
\tablefoottext{c}{Surface brightness detection limit in units of $10^{-17}$erg s$^{-1}$ cm$^{-2}$ arcsec$^{-2}$.}
\end{table}

\begin{table}
\begin{center}
\caption{Parameters of the putative H$\alpha$ emitters around TXS~2353-003 found in our observations.}
\begin{tabular}{c c c c c c c c c c}
\hline\hline 
ID  & RA        & Dec         & distance  & NB flux \tablefootmark{a} & BB flux \tablefootmark{a}     & BB-NB color &  EW$_{0}$ \tablefootmark{b}   & H$\alpha$~flux\tablefootmark{c} & SFR\tablefootmark{d}  \\
    &  (J2000)  & (J2000)     & [arcmin]  & [0.1$\mu$Jy] & [$\mu$Jy]   &       & [\AA]  & [10$^{-17}$~erg~s$^{-1}$~cm$^{-2}$] & [M$_{\odot}$ yr$^{-1}$]\\
\hline 
a   & 23:55:35.58 & -00:02:46.2 & 0.0       & 25.9 & 14.3 & 0.69  &  152.4 &     63.6 & N/A \\
b   & 23:55:37.95 & -00:01:36.8 & 1.29      & 16.2 & 11.3 & 0.95  &  100.2 &     32.1 & 35  \\
c   & 23:55:32.56 & -00:01:59.2 & 1.08      & 10.3 & 6.2  & 0.86  &  132.3 &     23.8 & 26  \\
d   & 23:55:37.64 & -00:03:23.9 & 0.81      &  8.3 & 5.4  & 0.76  &  112.7 &     18.7 & 20  \\
e   & 23:55:37.74 & -00:01:35.3 & 1.29      &  6.2 & 3.5  & 0.91  &  145.1 &     20.1 & 22  \\
f   & 23:55:31.92 & -00:01:33.8 & 1.51      &  6.7 & 3.5  & 1.01  &  166.8 &     17.1 & 19  \\
g   & 23:55:30.72 & -00:01:45.7 & 1.57      &  6.8 & 2.7  & 1.3   &  248.2 &     15.0 & 16  \\
\hline
\end{tabular}
\label{table:HalphaEmitters}
\end{center}
\tablefoottext{a}{Measured in 2\arcsec\ diameter apertures.}
\tablefoottext{b}{Rest-frame equivalent width.}
\tablefoottext{c}{Assuming that the color excess in the NB image is entirely from line emission.}
\tablefoottext{d}{Following \citet{Kennicutt1998}. We do not give a star formation rate (SFR) for the radio galaxy, because the line ratios are not consistent with photoionization from young stars.}

\end{table}

\newpage
\onecolumn
\appendix
\section{Candidate H$\alpha$ emitting galaxies around TXS~2353-003}

\begin{figure}
\centering
\includegraphics[width=0.9\textwidth]{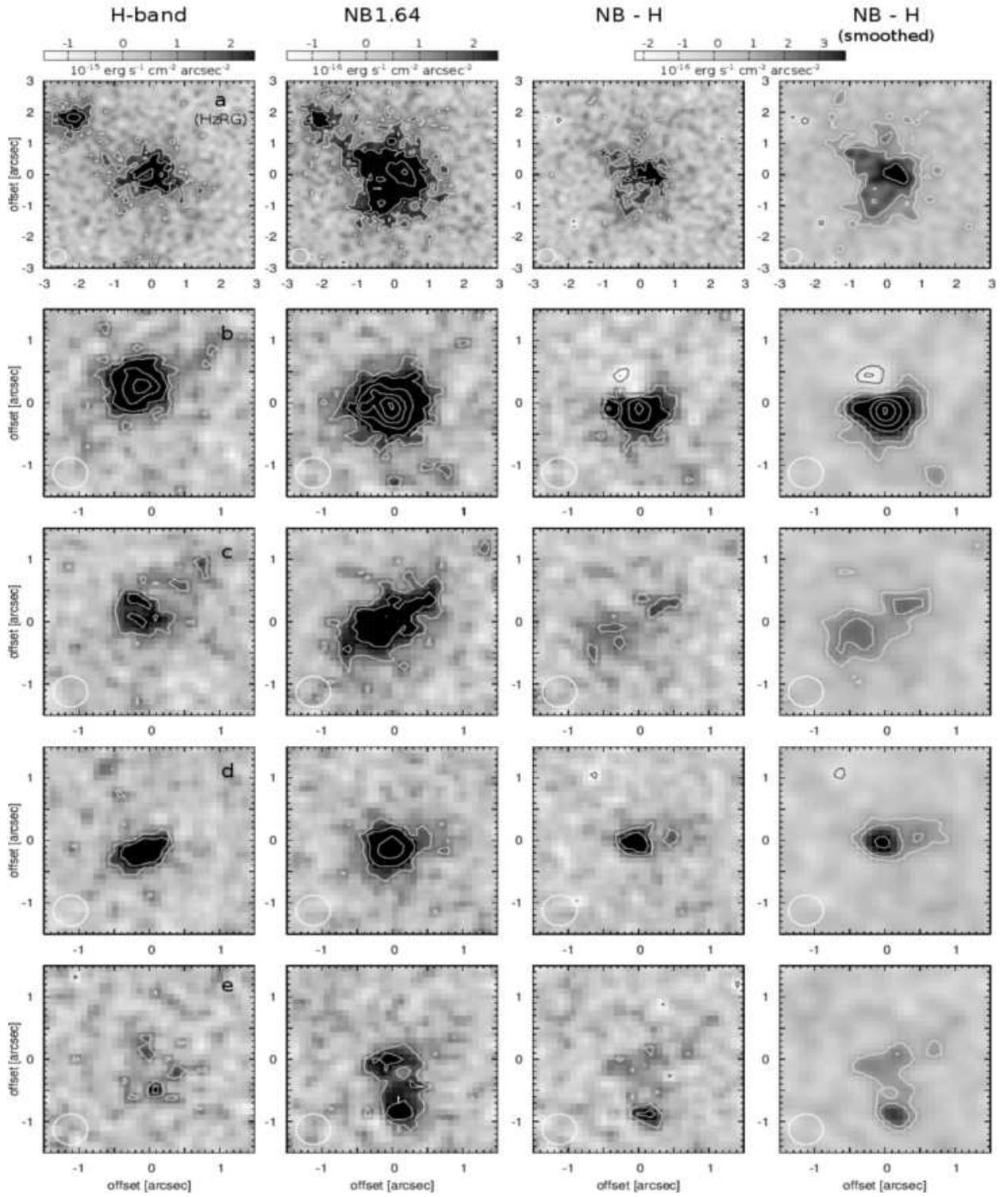}
\caption{Caption is under the last part of the image.}
\label{fig:thumbnailsHAE}
\end{figure}

\begin{figure}
\centering
\label{fig:thumbnailsHAE:b}
\includegraphics[width=0.9\textwidth]{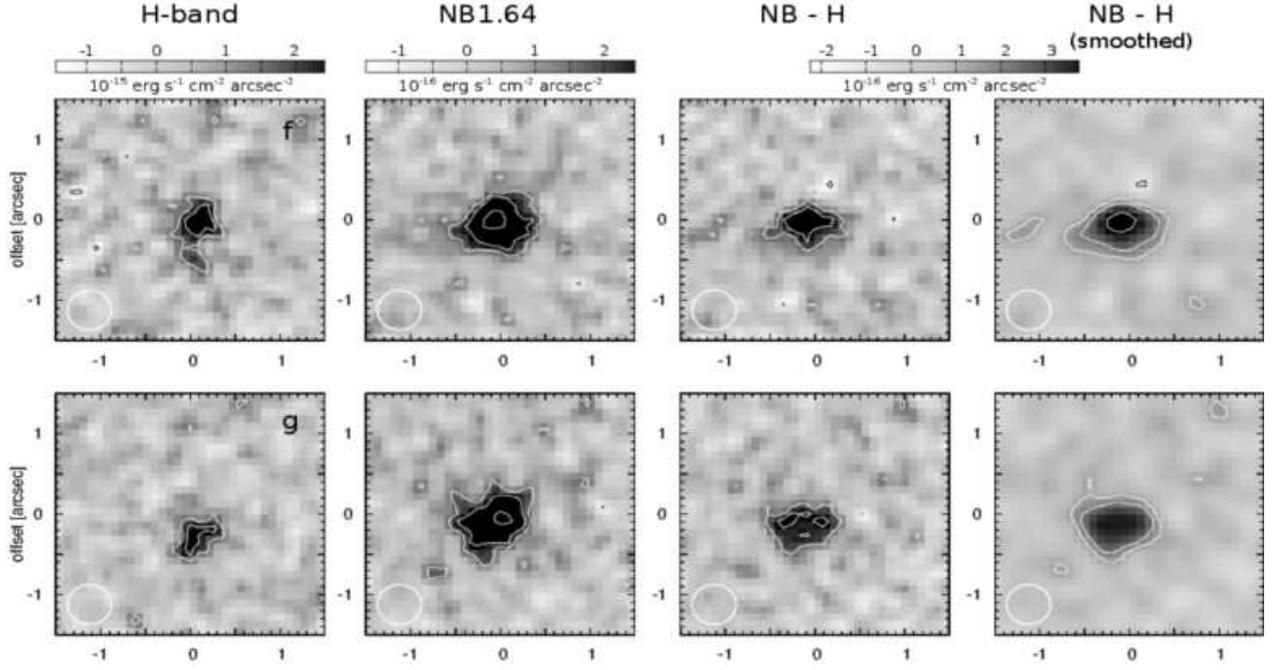}
\caption{ H$\alpha$ emitters around TXS~2353-003
  selected by the Bunker plot and confirmed when subtracting the
  continuum from the NB image. {\it Left column}: $H-$band images of
  the candidates. {\it Middle column}: NB1.64 images of the
  candidates. {\it Right columns}: Continuum-subtracted NB images of
  the candidates, the last column showing them smoothed with a
  Gaussian kernel of $\sigma = 1$~pixel.  The color palette spans 
  $-3 \sigma$ to $+5 \sigma$ in each case  in order to highlight the
  faint features of these HAE candidates. The contours give the
  morphology of the sources at larger surface brightnesses: white
  contours are at levels 3 ; 5 ; 10 ; 15 ; 20 $\sigma$ and the black
  contours are at levels $-10$ ; $-5$ ; $-3 \sigma$. The HzRG is 
  labeled $a$.} 
\end{figure}

\end{document}